\documentclass[10pt,journal,compsoc]{IEEEtran}
\usepackage{amsmath}
\usepackage{graphicx}
\usepackage{array}
\usepackage{epstopdf}
\usepackage{algorithm}
\usepackage{algorithmic}
\usepackage{epsfig}
\usepackage{subfigure}
\usepackage{multirow}

\ifCLASSOPTIONcompsoc
  \usepackage[nocompress]{cite}
\else
  \usepackage{cite}
\fi

\ifCLASSINFOpdf
\else
\fi

\begin{document}
%
\title{Scientific Article Recommendation: Exploiting Common Author Relations and Historical Preferences}

\author{Feng Xia,~\IEEEmembership{Senior~Member,~IEEE}, Haifeng Liu, Ivan Lee,~\IEEEmembership{Senior~Member,~IEEE}, \\
and Longbing Cao,~\IEEEmembership{Senior~Member,~IEEE}
\thanks{F. Xia and H. Liu are with the School of Software, Dalian University of Technology, Dalian 116620, China. E-mail: f.xia@ieee.org}
\thanks{I. Lee is with School of Information Technology and Mathematical Sciences, University of South Australia, Australia.}
\thanks{L. Cao is with the Advanced Analytics Institute, University of Technology, Sydney, Sydney, NSW 2007, Australia.}
}

\markboth{IEEE Transactions on Big Data}%
{Shell \MakeLowercase{\textit{et al.}}: Bare Demo of IEEEtran.cls for Computer Society Journals}

\IEEEtitleabstractindextext{
\begin{abstract}
Scientific article recommender systems are playing an increasingly important role for researchers in retrieving scientific articles of interest in the coming era of big scholarly data. Most existing studies have designed unified methods for all target researchers and hence the same algorithms are run to generate recommendations for all researchers no matter which situations they are in. However, different researchers may have their own features and there might be corresponding methods for them resulting in better recommendations. In this paper, we propose a novel recommendation method which incorporates information on common author relations between articles (i.e., two articles with the same author(s)). The rationale underlying our method is that researchers often search articles published by the same author(s). Since not all researchers have such author-based search patterns, we present two features, which are defined based on information about pairwise articles with common author relations and frequently appeared authors, to determine target researchers for recommendation. Extensive experiments we performed on a real-world dataset demonstrate that the defined features are effective to determine relevant target researchers and the proposed method generates more accurate recommendations for relevant researchers when compared to a Baseline method.
\end{abstract}

}

\maketitle

\IEEEdisplaynontitleabstractindextext

%
\IEEEpeerreviewmaketitle

\IEEEraisesectionheading{\section{Introduction}\label{sec:introduction}}
\IEEEPARstart{W}ith the rapid emergence of big scholarly data, tremendous growth of knowledge is now largely captured in digital form and archived all over the world. Archival materials are also currently being digitized and provided online to people for free or by paying a fee. Such situation creates the commonly known information overload problem especially in academia while bringing a significant advantage that allows people to easily access more knowledge. For example, a researcher in academia needs to find articles of interest to read for generating a research idea or citing an article related to the article he is writing, an author needs to submit his manuscript to a certain journal of which the topic is relevant to the manuscript, an editor needs to assign a manuscript to a reviewer who is an expert in the domain which the manuscript belongs to, or a researcher in a domain needs to collaborative with another researcher in another domain. These academic activities involve in an overwhelming number of articles, journals, reviewers, and researchers. Therefore, it is quite difficult for researchers to locate relevant articles, journals, reviewers, and researchers for the aforementioned purposes.

\begin{figure}
\centering
\epsfig{file=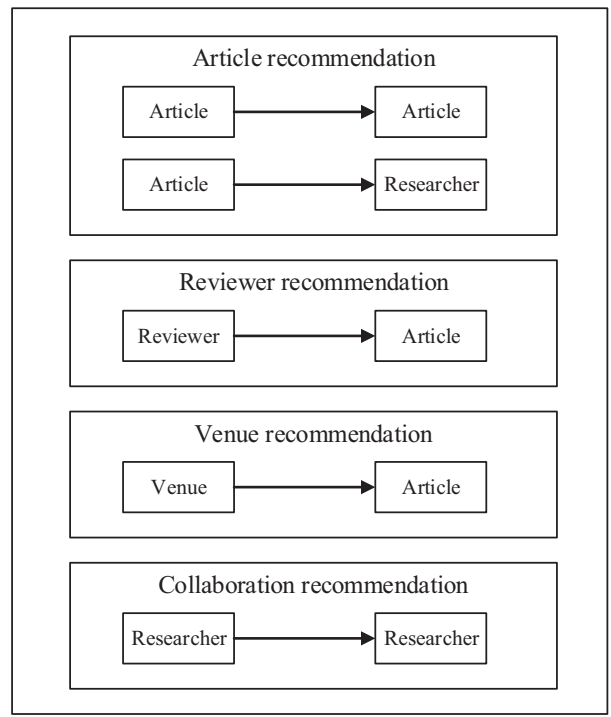,width=8cm}
\caption{Four academic recommendation tasks regarding to the entities in academia: researcher, article, venue, reviewer.}
\label{recommendation-task}
\end{figure}

Academic recommender systems aim to solve the information overload problem in big scholarly data such as finding relevant research paper, relevant publication venue, etc. Fig. \ref{recommendation-task} shows the corresponding recommendation tasks in above-mentioned scenarios, including (i) article recommendation \cite{gori2006research, tang2009discriminative, sun2013leveraging, haifeng2015car} for suggesting relevant articles to a researcher or an article for the purposes of reading or citation, (ii) reviewer recommendation \cite{kolasa2011survey, tayal2014new} for assigning a manuscript to the most appropriate reviewers (e.g., an expert in the same domain), (iii) venue recommendation \cite{yang2012venue, medvet2014publication} for suggesting a topic-relevant conference or journal to publish a new article, and (iv) collaboration recommendation \cite{tang2012cross, xia2014mvcwalker} for suggesting new partners to execute joint research (e.g., exploring cross-domain solution). There exist some interesting studies on these recommendation tasks. Gori and Pucci \cite{gori2006research} built a citation relation graph and employed a random walk algorithm to compute ranking scores of each possible citation. Tayal et al. \cite{tayal2014new} assigned relevant weights to various factors which affect the expertise of the reviewer to create a fuzzy set and then compute the expertise. Yang and Davison \cite{yang2012venue} extracted features related to writing-style information for computing similarity between articles and then applied traditional collaborative filtering to recommend a venue for submission. Xia et al. \cite{xia2014mvcwalker} considered three academic factors (i.e., co-author order, collaboration time, and number of collaboration) to define link importance, and then employed a random walk algorithm to compute rankings of potential collaborators.

In this paper, we focus on article-researcher recommendation, i.e., studying how to find articles of interest for target researchers in the context of big scholarly data. In the print age, researchers found articles of interest with the help of library catalogs. In recent years, web search tools employed by scientific digital libraries like IEEE Xplore, and literature search engines like Google Scholar, can retrieve a list of relevant articles in diverse technological fields using keyword-based queries. However, these search tools have several drawbacks as follows: (i) It is not enough to describe searchers' needs depending on only several limited keywords; (ii) The obtained results are the same for all searchers if only the keywords are the same; (iii) It is not feasible to search articles when a searcher has no ideas of what they are looking for. Article-researcher recommender systems aim to automatically suggest personalized articles of potential interest for different targets, thereby overcoming the problems stated above.

Existing studies \cite{sugiyama2010scholarly,sun2013leveraging} generally compute the content similarity between articles to find articles which are similar to the target's articles of interest, or compute the similarity between the target's profile and an new article's content to find matches. However, content extraction is not such simple because an article includes too many words. In this paper, we extract only author information to build relations between articles, i.e., common author relations. Then, these relations and researchers' historical preferences are used together to build a heterogenous graph for article ranking. The rationale of incorporating common author relations is that, the continuous development of internet technology enables researchers to easily build personal websites and share publications with others, which makes it more convenient to search articles published by the same author(s) for researchers who have such a search preference based on authors (we call it author-based search pattern). In addition, most studies ignore the fact that there exist different recommendation methods suitable for different targets. Therefore, we define features to find relevant target researchers who have author-based search patterns by analyzing information on common author relations existing in a researcher's historical preferences. In summary, we propose a novel Common Author relation-based REcommendation method (CARE) for specific target researchers with author-based search patterns.

Our main contributions in this work can be outlined as follows:

\setlength{\hangindent}{2em}
$\bullet$
We present two features including the ratio of pairwise articles with common author relations and the ratio of the most frequently appeared author, to help determine relevant researchers with author-based search patterns.

\setlength{\hangindent}{2em}
$\bullet$
We propose a novel recommendation method, which incorporates common author relations between articles to help generate better recommendations for relevant target researchers.

\setlength{\hangindent}{2em}
$\bullet$
We conduct relevant experiments using a real-world dataset CiteULike to evaluate the impacts of the defined features and the performance of the proposed method. In addition, other two features have also been defined and proved to be not effective for determining suitable targets.

The rest of the paper is organized as follows. Section 2 reviews related work on article recommendation. Section 3 presents our problem definition. Section 4 introduces the details of our proposed method. Section 5 describes our experimental setup and discusses our results in detail. Section 6 finally concludes the paper.

\section{Related Work}
Recommender systems aim to automatically suggest items of potential interest to users. As well-known effective tools for solving information overload problems, recommender systems have been successfully applied in multiple domains including traffic \cite{qu2014cost}, movies \cite{tang2013cross,shi2013mining}, music \cite{li2007probabilistic,kaminskas2013location}, news \cite{garcin2013personalized}, e-commerce \cite{wang2012strategy,wang2013opportunity}, e-learning \cite{bobadilla2009collaborative,hsu2008proposing}, and so on. As aforementioned, with the rapid development of information technology and ever-growing amounts of scholarly data, it is becoming increasingly popular and challenging to apply recommendation techniques in academia. In this section, we focus on reviewing related work on article recommendation.
\subsection{Article-Article Recommendation}
Article-article recommendation, i.e., citation recommendation, includes global citation recommendation \cite{strohman2007recommending,gori2006research,bethard2010should,nallapati2008joint,meng2013unified,ren2014cluscite,liu2014meta} and local citation recommendation \cite{tang2009discriminative,lu2011recommending,huang2012recommending,tang2014cross,huang2015neural}. Global citation recommendation aims to recommend a list of citations for a given query article. Strohman et al. \cite{strohman2007recommending} linearly combined text features and citation graph features to measure the relevance between articles. They conducted relevant experiments with their proposed citation recommender system and concluded that similarity between bibliographies of articles and Katz distance are the most important features. Gori and Pucci \cite{gori2006research} used citation relations between articles to built a citation graph and applied a random walk algorithm in the graph to compute ranking scores of each article as a reference of a target article. Bethard and Jurafsky \cite{bethard2010should} incorporated a wide variety of features (including author impact, author citation habits, citation count, and publication ages) to build a retrieval model for literature search. After a training process, the model took abstract of an article as input to produce relevant reference lists. Nallapati et al. \cite{nallapati2008joint} jointly modeled the text and citation relationship under a framework of topic model. They introduced a model Pair-Link-LDA which models the presence or absence of a link between pairwise articles and does not scale to large digital libraries. They also introduced another model called Link-PLSA-LDA which models citations as a sample from a probability distribution associated with a topic. Meng et al. \cite{meng2013unified} incorporated various types of information including content, authorship, and collaboration network to build a unified graph-based model for personal global citation recommendation. Ren et al. \cite{ren2014cluscite} proposed to cluster article citations into interest groups to determine the significance of different structural relevance features for each group while deriving an article's relative authority within each group. Liu et al. \cite{liu2014meta} employed the pseudo relevance feedback (PRF) algorithm to determine important nodes like authors and venues on a heterogeneous bibliographic graph. Then, a random walk algorithm was run to compute the ranking scores of an article.

On the other hand, local citation recommendation aims to recommend citations for a given specific context such as a sentence in a paper. Tang and Zhang \cite{tang2009discriminative} formally defined the problem of topic-based citation recommendation and proposed to model article contents and citation relationships using a two-layer restricted Boltzmann machine. For a given context, they calculated the probability of each article being the reference based on the model. Lu et al. \cite{lu2011recommending} proposed to recommend citations using a translation model which is originally used in translating text in one language to another. They assumed that the languages used in citation contexts and article's content are different and translated one word in context to one word in citation. Based on the probability of translating one word to another, relevant articles were recommended to a citation context. Huang et al. \cite{huang2012recommending} regarded an article as new 'words' in another language and employed a translation model for estimating the probability of citing an article given a citation context. Tang et al. \cite{tang2014cross} proposed a cross-language context-aware citation recommendation method for the purpose of recommending English citations for a given context of the place where a citation should be made in a Chinese article. Huang et al. \cite{huang2015neural} proposed a novel neural probabilistic model which jointly learns the semantic representations of citation contexts and cited articles and then estimated the probability of citing an article by a neural network.

\subsection{Article-Researcher Recommendation}
Article-researcher recommendation is our focus in this paper. Most existing studies compute similarities among researchers and articles based on articles' contents \cite{sugiyama2010scholarly, sugiyama2014comprehensive, wang2011collaborative, nascimento2011source, jiang2012recommending, tian2013recommending, sun2013leveraging, pera2011personalized, pera2014exploiting} or tags in social tagging systems \cite{sun2013leveraging, pera2011personalized, pera2014exploiting, xia2014folksonomy} and then apply traditional collaborative filtering to generate recommendations. Sugiyama and Kan \cite{sugiyama2010scholarly} examined the effect of modeling a researcher's past works in scientific article recommendation to the researcher. A researcher's profile was derived from his past works and other works which are the references or citations of those works. Apart from previous explicit citations, Sugiyama and Kan \cite{sugiyama2014comprehensive} additionally took into account implicit citations. Potential citation articles were discovered using collaborative filtering and then combined with previous information to enhance the profiles of candidate articles and researchers. Finally, the two types of profiles were compared to compute similarity as their cosine measure. Wang and Blei \cite{wang2011collaborative} proposed a collaborative topic regression model for article recommendation, where each user was represented with interest's distribution and each article was described using content-based item topic distribution. Nascimento et al. \cite{nascimento2011source} proposed a framework to generate structured search queries and obtained candidate articles using existing web information sources. Then, they computed the content-based similarity for ranking candidate articles. Jiang et al. \cite{jiang2012recommending} employed a concept-based topic model to compute the problem-based similarity and solution-based similarity between a known article of interest and an unknown article. Tian and Jing \cite{tian2013recommending} employed LDA (Latent Dirichlet Allocation) model to obtain each article's representation based on content and computed the similarity between articles for determining their associations. Sun et al. \cite{sun2013leveraging} exploited semantic content and heterogeneous connections (i.e., social connection, behavioral connection, and semantic connection) to build two kinds of profiles of researchers and then computed researcher-article similarities and researcher-researcher similarities. Using a Social-Union method \cite{symeonidis2011product}, the score of a target researcher on an article is defined by that of his nearest neighbors. Finally, it is fused with the results obtained based on researcher-article similarity computation to compute the final ranking scores. Pera and Ng \cite{pera2011personalized,pera2014exploiting} proposed a personalized recommender for scientific articles, called PReSA. Given a target publication $P$ which a user $U$ expressed interest in, PReSA computes the three similarities between $P$ and each candidate publication $CP$ which is from the ones in the personal libraries of $U$'s connections: a) tag similarity, b) title similarity, and c) abstract similarity. Then, these similarities and popularity of publications were fused to calculate the ranking score of $CP$ by employing a weight linear combination strategy. Finally, top-N publications were recommended to $U$. Xia et al. \cite{xia2014folksonomy} built a active participant's profile and each group's profile based on tags annotated by participants. Then, their similarity was computed to recommend the active participant's to other participants in groups with higher similarities.

In this paper, we utilize only information on articles' authors to build common author relations between articles. Compared to analysis on content and tags, our work is simpler and more time-saving, because the number of terms in content and tags is enormous and there exist lots of irrelevant terms. In addition, these studies do not take into account specific target researchers suitable for their recommendation methods. We assume that, since our proposed method (CARE) incorporates common author relations, only a part of researchers can be selected as targets for high-quality recommendation. Accordingly, we define features to determine such researchers.

\begin{figure}
\centering
\epsfig{file=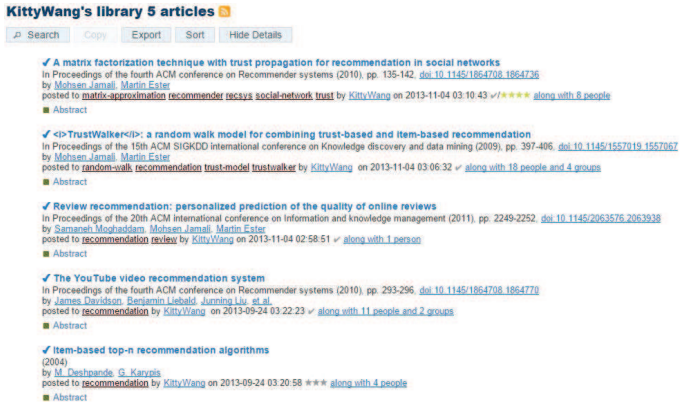,width=9cm}
\caption{An example of article library.}
\label{citeulike}
\end{figure}

\begin{figure}
\centering
\epsfig{file=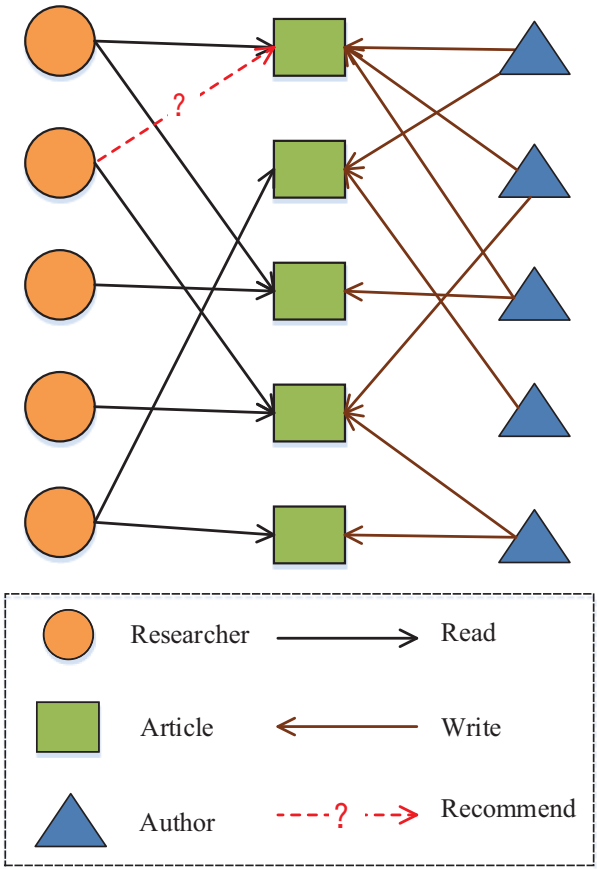,width=8cm}
\caption{A example of recommendation scenario including three entities (researcher, article, and author) and two relations (reading and writing).}
\label{scenario}
\end{figure}

\section{Problem Statement}
In academic social tagging websites such as CiteULike, each of the registered users is generally a researcher. When a researcher is interested in an article, he will post it into his article library, read it extensively and then tag the article with one or more special keywords. As shown in Fig. \ref{citeulike}, five articles have been included in KittyWang's library and each of them is given many different tags by the researcher. The researcher's historical preference is represented by the set of articles that interest him in his library. In this paper, our scientific article recommendation method aims to study how to automatically find the most possibly-preferred articles which will be posted into a target researcher's library.

Fig. \ref{scenario} shows our recommendation scenario. The scenario includes three objects: researchers, articles, and authors. Although there are possibly overlaps between researchers and authors, this situation is not taken into consideration due to the facts that the CiteULike website does not provide enough information such as email address to determine each registered researcher's identity (i.e., whether he is an author of a certain article in article library) and the recommendation targets are the registered researchers rather than the authors. Additionally, there exist two kinds of links among the three objects. The first kind of link represents that a researcher, who is a registered user in CiteULike website, has read one or more articles he is interested in. The second kind of link represents that an article is written and published by one or more authors. Traditional collaborative filtering methods utilize the first kind of link to generate recommendations. The rational underlying these methods is that, two researchers who are interested in the same articles are similar and then the tastes of similar researchers are used to predict those of target researchers. Generally, the second kind of link is ignored in these collaborative filtering methods. However, these additional information may influence recommendation quality due to the fact that researchers often focus the same authors' publications when they find these authors' work relevant to theirs. As a result, it is very necessary to incorporate the second kind of link to propose a novel recommendation method. This is the first problem we address in this paper.

Researchers search articles published by the same authors to find articles they are interested in. We call it author-based search pattern. Actually, not all researchers employ this pattern. Some of them are likely to search citations or only focus on articles' titles. Therefore, the second problem we are aiming to solve is how to find target researchers with author-based search patterns. Then, the information on authors can be incorporated to help generate better recommendations for these targets.

\begin{figure}
\centering
\epsfig{file=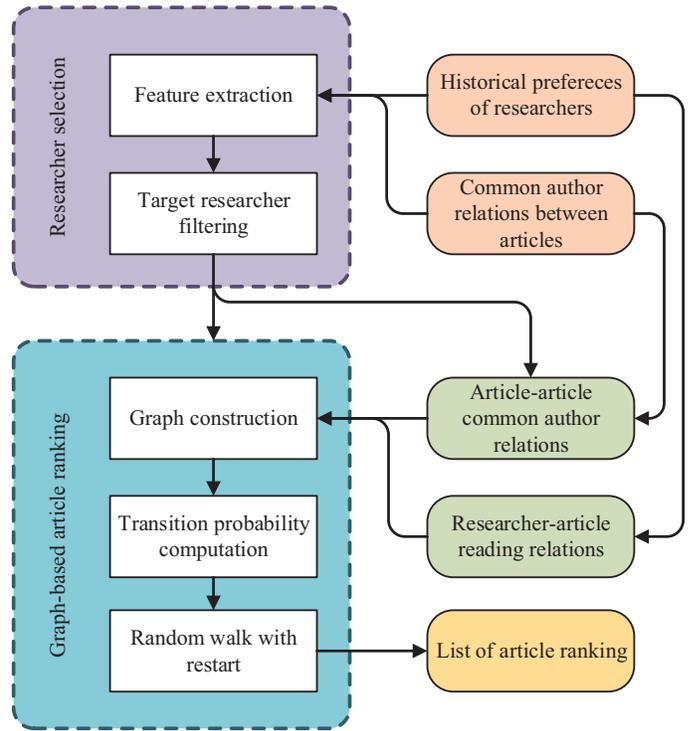,width=9cm}
\caption{The architecture of CARE.}
\label{architecture}
\end{figure}

\section{Design of CARE}
\subsection{Overview}
Our CARE method is inspired by two important facts: (i) researchers generally search articles written by the same authors; (ii) not all researchers have such an author-based search patterns. Fig. \ref{architecture} shows the architecture of CARE, which mainly includes two components: (i)researcher selection module and (ii) graph-based article ranking module. The first component is responsible for extracting relevant features from researchers' historical preferences and then selecting researchers with author-based search patterns as recommendation targets. The second component is responsible for incorporating common author relations to build a graph and generating article ranking list through a graph-based random walk algorithm. In the domain of recommender systems, random walk-based ranking is a classical technique for recommendation. Based on the technique, many researchers \cite{gori2007itemrank} have successfully applied it to various recommendation scenarios. Next, we will introduce the two components in detail.

\subsection{Target researcher selection}
\begin{figure*}
\centering
\epsfig{file=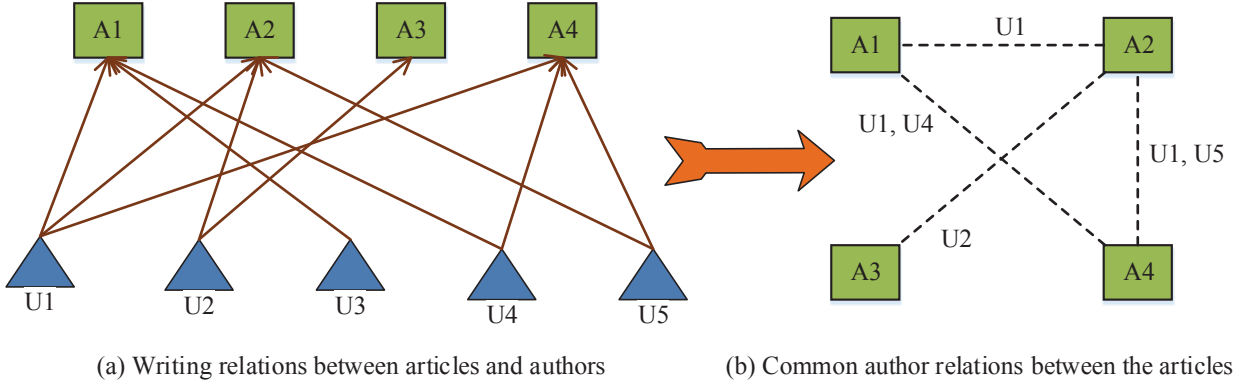,width=0.9\textwidth}
\caption{An example scenario for a researcher.}
\label{ComputationExample}
\end{figure*}

For researchers who find articles of interest by searching article written by the same authors, in their online article libraries, possibly there are lots of articles which are mainly written by one or several authors. Therefore, we define two features which are relevant to common authors between any two articles to help determine target researchers:

$\bullet$
$FE1$, is the ratio of the total number of pairwise articles with common author relations to the total number of all pairwise articles for a researcher.

$\bullet$
$FE2$, is the ratio of the occurrence number of the most frequently appeared author in articles to the total number of articles for a researcher.

For a researcher, when $FE1$ or $FE2$  is larger than a given threshold, this researcher will be considered to have an author-based search pattern and will be regarded as a target which is suitable for the next ranking component of our CARE method.

We use Fig. \ref{ComputationExample} as an example scenario for illustrating the computation process of the above two features. Fig \ref{ComputationExample}(a) shows the writing relations between article and author for a target researcher $X$, where each edge links an article to one of all its authors and researcher $X$ expressed interest to all articles in the figure. For example, article $A1$ is linked to its three authors $U1$, $U3$, and $U4$ through three different edges respectively. We consider two articles are related if they are linked to the same author(s) in Fig. \ref{ComputationExample}(a). In this way, we convert the writing relation graph into a common author relation graph, as shown in Fig. \ref{ComputationExample}(b), where two articles are linked to each other if they have common author(s). From Fig. \ref{ComputationExample}(b), we can easily obtain the number of pairwise articles and it is equal to 4. For these articles, the number of all possible relations between articles is equal to $C_2^4=6$. As a result, $FE1$ is equal to $4/6=0.67$. In addition, $U1$ is the most frequently occurred author and its occurrence number is 3. Then, $FE2$ is equal to $3/4=0.75$. If the thresholds of $FE1$ and $FE2$ are set to 0.2 and 0.3 respectively, then $0.67>0.2$ and $0.75>0.3$. Therefore, the researcher $X$ is a relevant target suitable for CARE method. There may be other features to determine relevant target researchers, but in this paper we consider the above two features and conduct experiments to verify their effectiveness in Section 5.


\subsection{Graph-based Article Ranking}
\subsubsection{Graph Construction}
As aforementioned, in field of academic recommendation, there are many entities such as researchers, articles, conferences, journals, and so on. In this paper, we consider some of them and then design a method for recommending scientific articles. Scientific article recommender systems include a set of $n$ researchers $R=\{R_1,\cdots,R_n\}$ and a set of $m$ articles $A=\{A_1,\cdots,A_m\}$. Based on researchers' historical preferences, we can give the pairwise reading relations between researchers and articles, denoted as $W_{RA}=W_{RA}^{n\times m}$ with $W_{RA}(i,j)$ and $W_{AR}=W_{AR}^{m\times n}$ with $W_{AR}(j,i)$, indicating whether a researcher $R_i$ has read and expressed interest in an article $A_j$, as shown in Equation (\ref{RA-Matrix}). As we consider undirected relations in our method, $W_{RA}(i,j)$ is equal to $W_{AR}(j,i)$. As stated in the previous section, we convert the writing relations between articles and authors into common author relations between articles. Then, we also give the pairwise relations between articles, denoted as $W_{AA}=W_{AA}^{m\times m}$ with $W_{AA}(i,j)$ indicating whether there is/are common author(s) between the two article $A_i$ and $A_j$, as shown in Equation (\ref{AA-Matrix}). Likewise, we employ undirected common author relations. Therefore, $W_{AA}(i,j)$ is equal to $W_{AA}(j,i)$. In addition, as there is no consideration on relations between researchers, we denote $W_{RR}=[0]$.
\begin{equation}
W_{RA}(i,j)=\left\{
\begin{aligned}
&1   \hspace{5mm} \text{\textbf{if} $R_i$ expressed interest in $A_j$} \\
&0   \hspace{5mm} \text{\textbf{otherwise}}
\end{aligned}
\right.
\label{RA-Matrix}
\end{equation}
\begin{equation}
W_{AA}(i,j)=\left\{
\begin{aligned}
&1   \hspace{5mm} \text{\textbf{if} there is/are common author(s)} \\
&    \hspace{10mm} \text{ between articles $A_i$ and $A_j$} \\
&0   \hspace{5mm} \text{\textbf{otherwise}}
\end{aligned}
\right.
\label{AA-Matrix}
\end{equation}

\begin{figure}
\centering
\epsfig{file=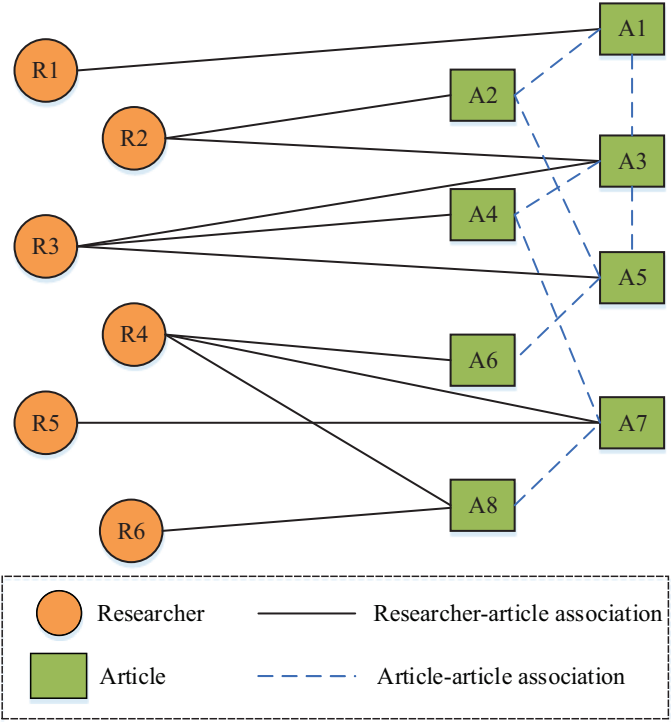,width=9cm}
\caption{An example graph for article ranking.}
\label{graph}
\end{figure}

Based on the above two relation matrices, we construct a graph for applying a random walk-based article ranking algorithm, as shown in Fig. \ref{graph}. Let $G=(V_R\bigcup V_A,E_{RA}\bigcup E_{AA})$, where $E_{RA}\subseteq V_R\times V_A$ and $E_{AA}\subseteq V_A\times V_A$. $V_R$ and $V_A$ indicate the set of researcher vertices and the set of article vertices, respectively. $E_{RA}$ and $E_{AA}$ describe the set of reading relations between researchers and articles and the set of common author relations between articles. An edge linking a researcher $i$ to an article $j$ exists in the graph if $W_{RA}(i,j)$ or $W_{AR}(j,i)$ is equal to 1. Similarly, an edge linking an article $i$ to another article $j$ exists in the graph if $W_{AA}(i,j)$ or $W_{AA}(j,i)$ is equal to 1.

\subsubsection{Transition Probability Computation}

A random walk in the graph is actually a transition from a vertex to another vertex. Therefore, we subsequently utilize the above three matrices to build a transition matrix, of which each element represents the transition probability between two corresponding vertices (article to article, article to researcher, and researcher to researcher). The computation process is as follows. When a random walk starts with a researcher vertex, the transition probability of moving to another researcher vertex is
\begin{equation}
\begin{aligned}
T_{RR}(i,j)=0
\end{aligned}
\end{equation}
and the transition probability of moving to an article vertex is
\begin{equation}
\begin{aligned}
T_{RA}(i,j)=\frac {W_{RA}(i,j)}{\sum_{k}W_{RA}(i,k)}
\end{aligned}
\end{equation}
Additionaly, when a random walk starts with an article vertex, the transition probability of moving to another article vertex is
\begin{equation}
\begin{aligned}
T_{AA}(i,j)=\frac {W_{AA}(i,j)}{\sum_{k1}W_{AR}(i,k1)+\sum_{k2}{W_{AA}(i,k2)}}
\end{aligned}
\end{equation}
and the transition probability of moving to a researcher vertex is
\begin{equation}
\begin{aligned}
T_{AR}(i,j)=\frac {W_{AR}(i,j)}{\sum_{k1}W_{AR}(i,k1)+\sum_{k2}{W_{AA}(i,k2)}}
\end{aligned}
\end{equation}
The transition probability matrix is
\begin{equation}
T=\left[
\begin{array}{cc}
T_{RR} & T_{RA}  \\
T_{AR} & T_{AA}
\end{array}
\right]
\end{equation}

Note that, in the above computation process of transition probability matrix, for each vertex, we assign equal values to all its neighbor vertices no matter what kind of vertex (researcher and article) the neighbor is. Specially, a vertex moves to any one of its neighbor vertices with the same probability even though these neighbors are different types of vertices.
\subsubsection{Random Walk with Restart}
After obtaining the transition probability matrix, a random walk with restart method is employed to compute articles' rankings. Generally, the algorithm finds articles of interest based on the meta path: researcher-article-researcher-article. This means, a researcher is likely to be interested in an article which another researcher who has similar historical preferences expressed interest to. We incorporate common author relations between articles and then add another meta path: researcher-article-article. This means, a researcher is likely to be interested in an article which is similar to another article which another researcher has expressed interest to. Our algorithm considers the two meta paths. Starting from a source vertex $v_0$ (target researcher), we perform random walk with restart in the graph built in previous section. After walking to any vertex $v_x$, we continue the next random walk with probability $\alpha$ and walk to another vertex $v_y$ which links to $v_x$ with transition probability $T(v_x,v_y)$. With probability $1-\alpha$, we return to source vertex $v_0$. Algorithm \ref{random-walk} shows the process of graph-based article ranking. In this algorithm, a list of article rankings for target researcher are computed, and top-N articles which the researcher have not expressed interest in before, will be put in the recommendation list for a target researcher.
\begin{algorithm}[htb]
\caption{Graph-based article ranking.}
\begin{algorithmic}[1]
\REQUIRE ~~\\  
  Graph, $G$;\\
  Random walk probability, $\alpha$;\\
  Target researcher vertex, $v_0$;\\
  Maximum step length of iteration, $maxStep$;\\
  Transition probability matrix, $T$;
\ENSURE ~~\\  
  Ranking scores of all article vertices, $ScoreArticle(1:m)$; // $m$ article vertices
\STATE Define ranking scores of all vertices, $ScoreAll(1:n+m)$; // $n+m$ vertices
\FOR{each $v\in {V_R\bigcup V_A}$}
  \STATE $ScoreAll(v)=0$;        //initial ranking scores are 0
\ENDFOR
\STATE $ScoreAll(v_0)=1$;     
\FOR{$step=0$; $step<maxStep$; $step++$}
  \FOR{each $v\in {V_R\bigcup V_A}$}
    \STATE $tmpScore(v)=0$;  //initial values are 0
  \ENDFOR
  \FOR{each $v_x\in {V_R\bigcup V_A}$}
     \FOR{each $v_y\in {V_R\bigcup V_A}$}
        \STATE $tmpScore(v_y)=\alpha \times ScoreAll(v_x)\times T(v_x,v_y)+tmpScore(v_y)$;
     \ENDFOR
     \IF{$v_x==v_0$}
        \STATE $tmpScore(v_x)=tmpScore(v_x)+1-\alpha$;
     \ENDIF
  \ENDFOR
  \STATE $ScoreAll=tmpScore$;
\ENDFOR
\STATE $ScoreArticle(1:m)=ScoreAll(n+1:n+m)$; // select ranking scores of article vertices
\RETURN $ScoreArticle(1:m)$;
\end{algorithmic}
\label{random-walk}
\end{algorithm}

\section{EXPERIMENTS}
\subsection{Dataset}
\begin{table}
\centering
\caption{Data statistics.}
\begin{tabular}{|c|c|} \hline
Number of researchers& 5550 \\ \hline
Number of articles& 15439  \\ \hline
Number of researcher-article reading relations& 200251  \\ \hline
Sparsity of researcher-article reading relations& 0.9977  \\ \hline
Number of article-article common author relations& 18646  \\ \hline
Sparsity of article-article common author relations& 0.9998  \\ \hline
\end{tabular}
\label{statistics}
\end{table}

CiteULike is a free web-based tool to help scientists, researchers, and academics store, organize, share, and discover links to academic research papers. With more than 3.5 million papers currently bookmarked and over 900000 visitors per month, CiteULike has grown to be one of the biggest and most popular social reference management websites by helping users streamline their process of storing and managing academic references. Emamy and Cameron \cite{emamy2007citeulike} have provided detail description on CiteULike. We used the version of CiteULike dataset collected by Wang et al. \cite{wang2011collaborative} in our experiments. This dataset includes all registered users' (researchers) historical preferences, i.e., articles in each user's library, and articles' contents. Note that there is no author information in the original dataset. We designed a web crawler to collect each article's author information from CiteULike website. Then, we compared pairwise articles' authors to determine their common author relations. To avoid the situation that some authors' names are the same, two articles are considered to be relevant if they have at least two same authors. Although there possibly exist his own article(s) in a researcher's library, this situation can be ignored due to the following facts: (i) CiteULike has not provided enough information on the registered researchers' identities such as email address, so there is no way to determine whether an article in a researcher's article library belongs to him (i.e., its author); (ii) The registered researchers generally put other researchers' articles of interest into his library. The original dataset includes 5551 researchers and 16980 articles. We removed articles with less than 5 researchers who express interests in them. The distribution of the preprocessed dataset is shown in Table \ref{statistics}. Similar to most datasets for evaluating recommendation methods, this one has the characteristic that the researcher-article and article-article relations are very sparse, i.e. data sparsity. The sparsity indicates the ratio of the difference between numbers of all possible relations and existing common author relations to the number of all possible relations. Therefore, based on the spare data, if a novel recommendation method can be designed to improve recommendation quality, to some extent, the challenge of data sparsity will be solved.
\begin{figure*}
\centering
\subfigure[Precision]{
\label{alpha-precision}
\includegraphics[width=0.325\textwidth]{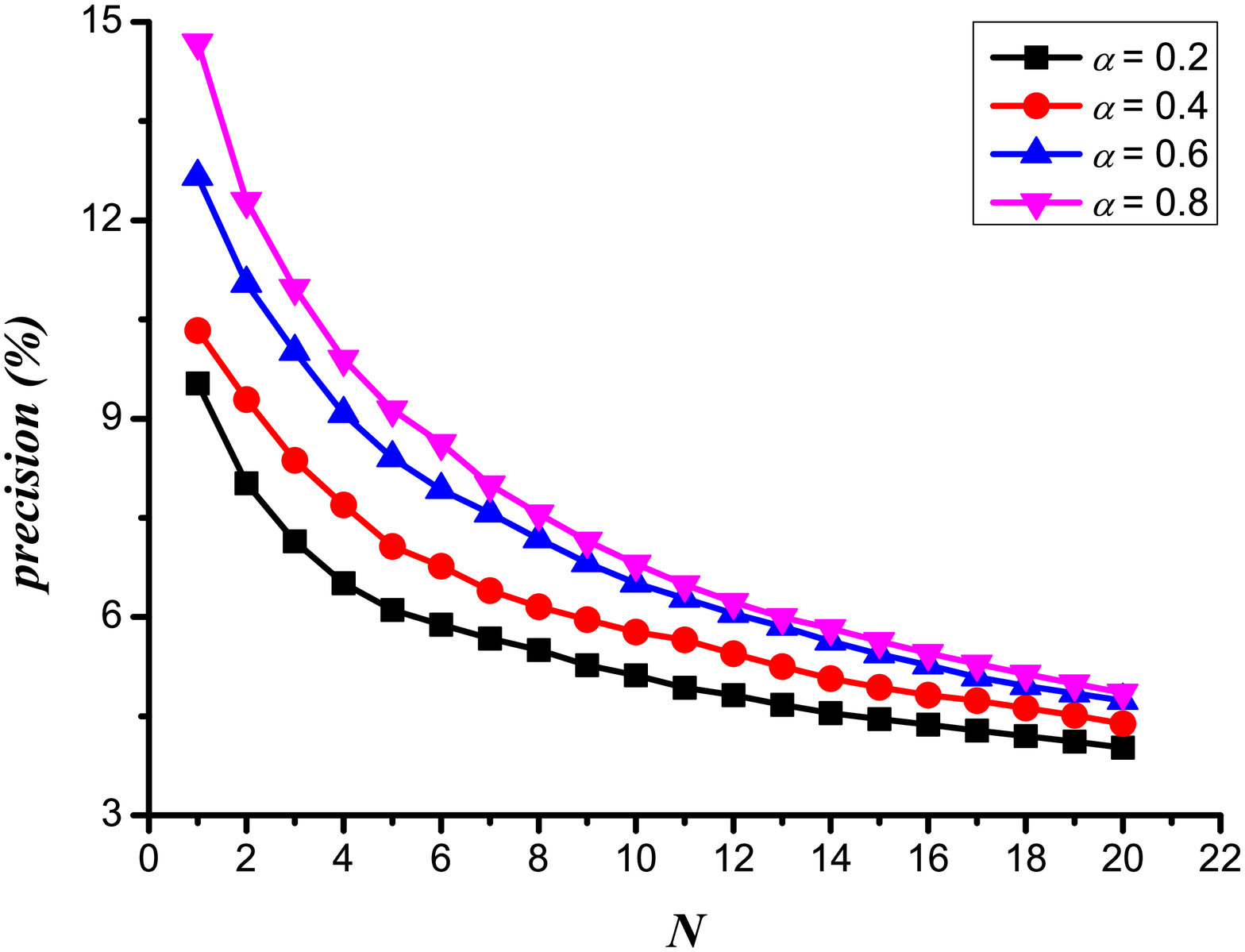}}
\subfigure[Recall]{
\label{alpha-recall}
\includegraphics[width=0.325\textwidth]{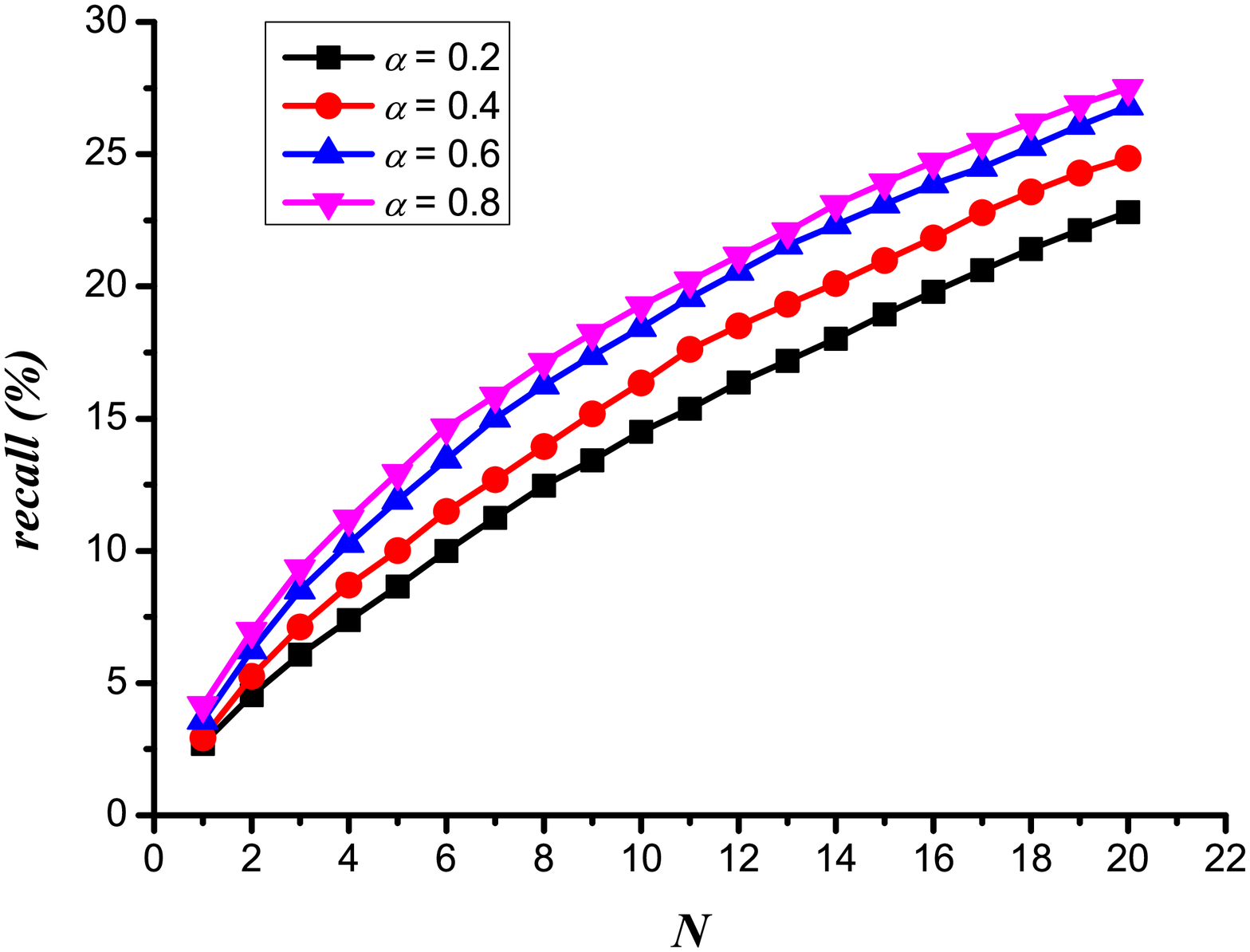}}
\subfigure[F1]{
\label{alpha-f1}
\includegraphics[width=0.325\textwidth]{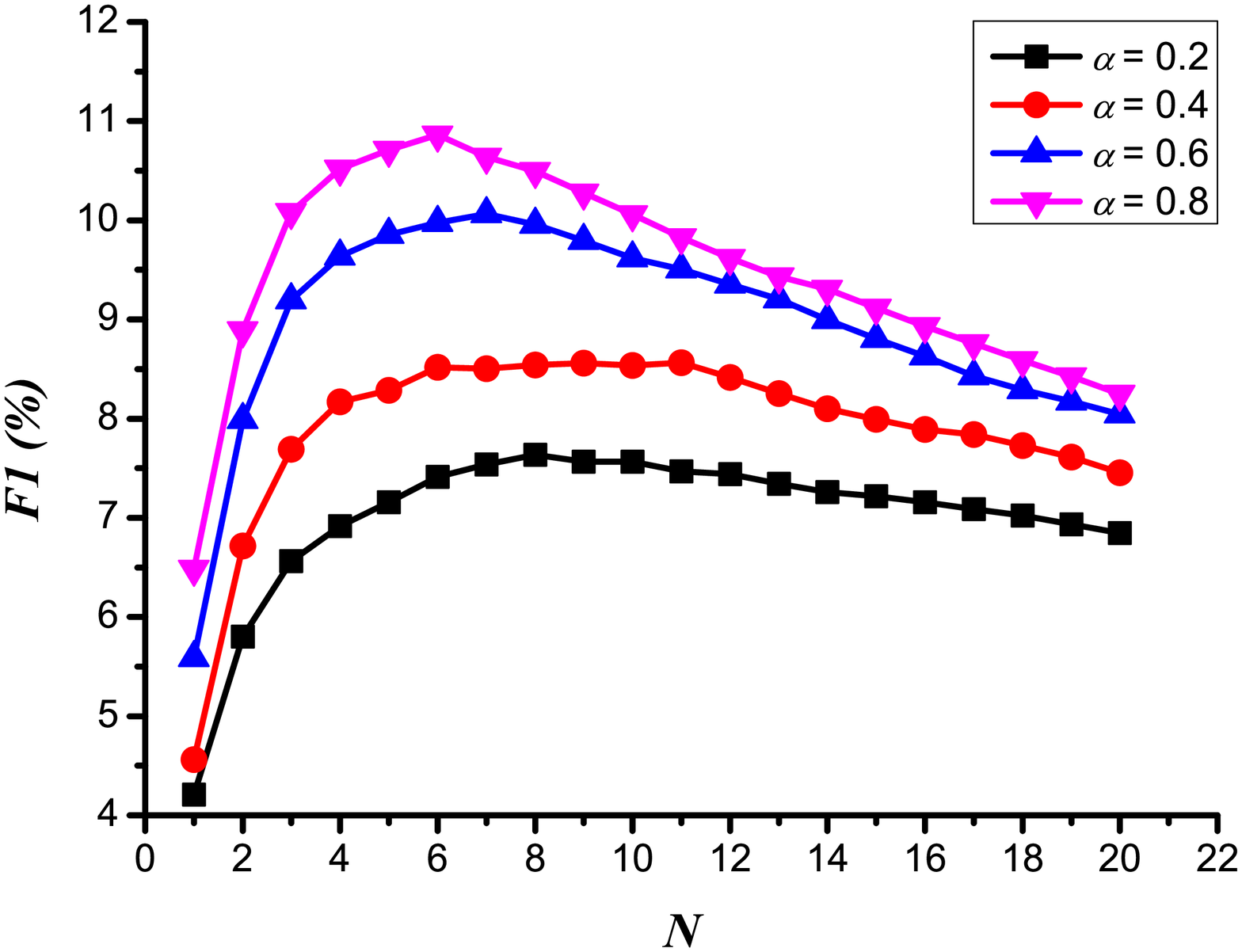}}
\caption{Precision, recall, and F1 of CARE for different walking probability $\alpha$.}
\label{alpha-impact}
\end{figure*}

\begin{figure*}
\centering
\subfigure[Precision]{
\label{precision}
\includegraphics[width=0.325\textwidth]{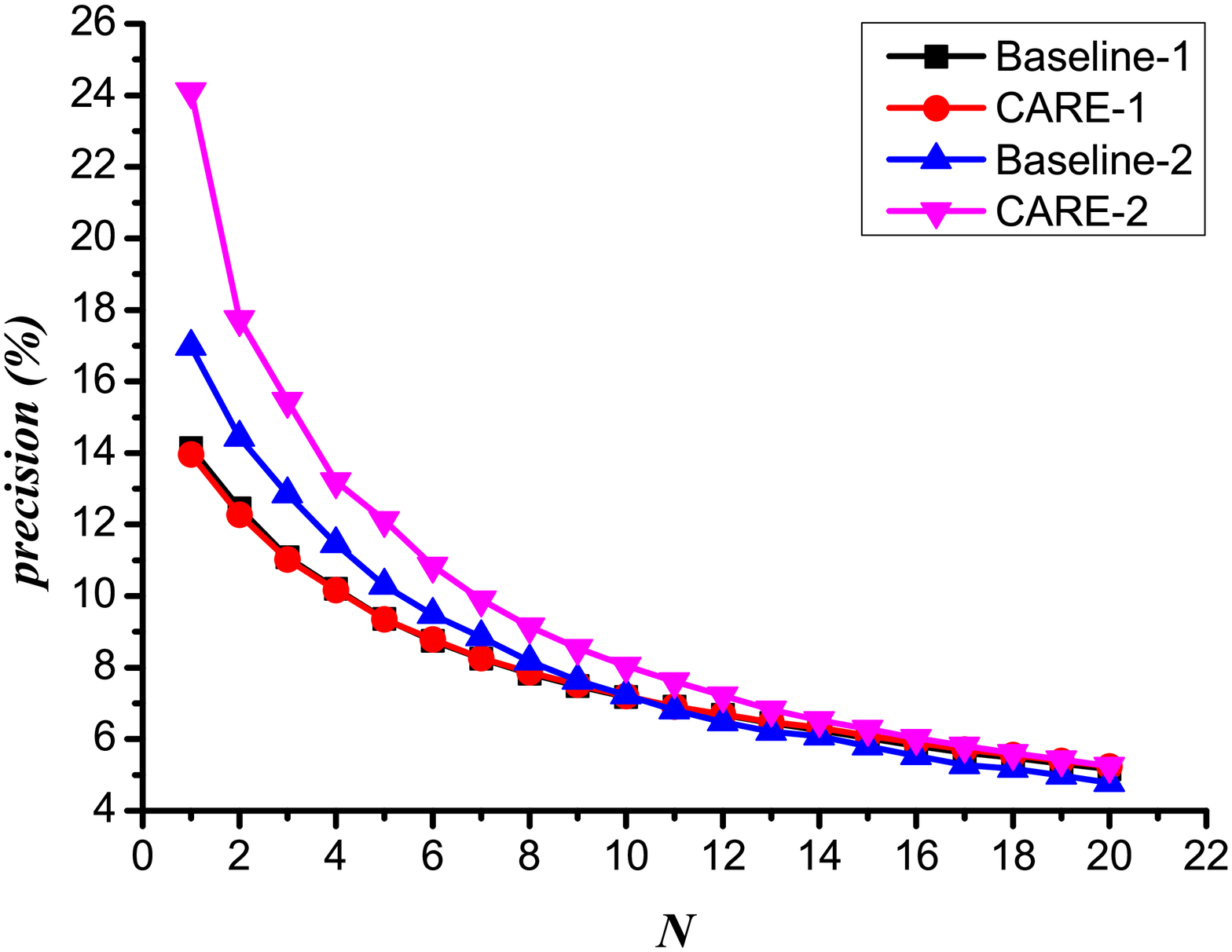}}
\subfigure[Recall]{
\label{recall}
\includegraphics[width=0.325\textwidth]{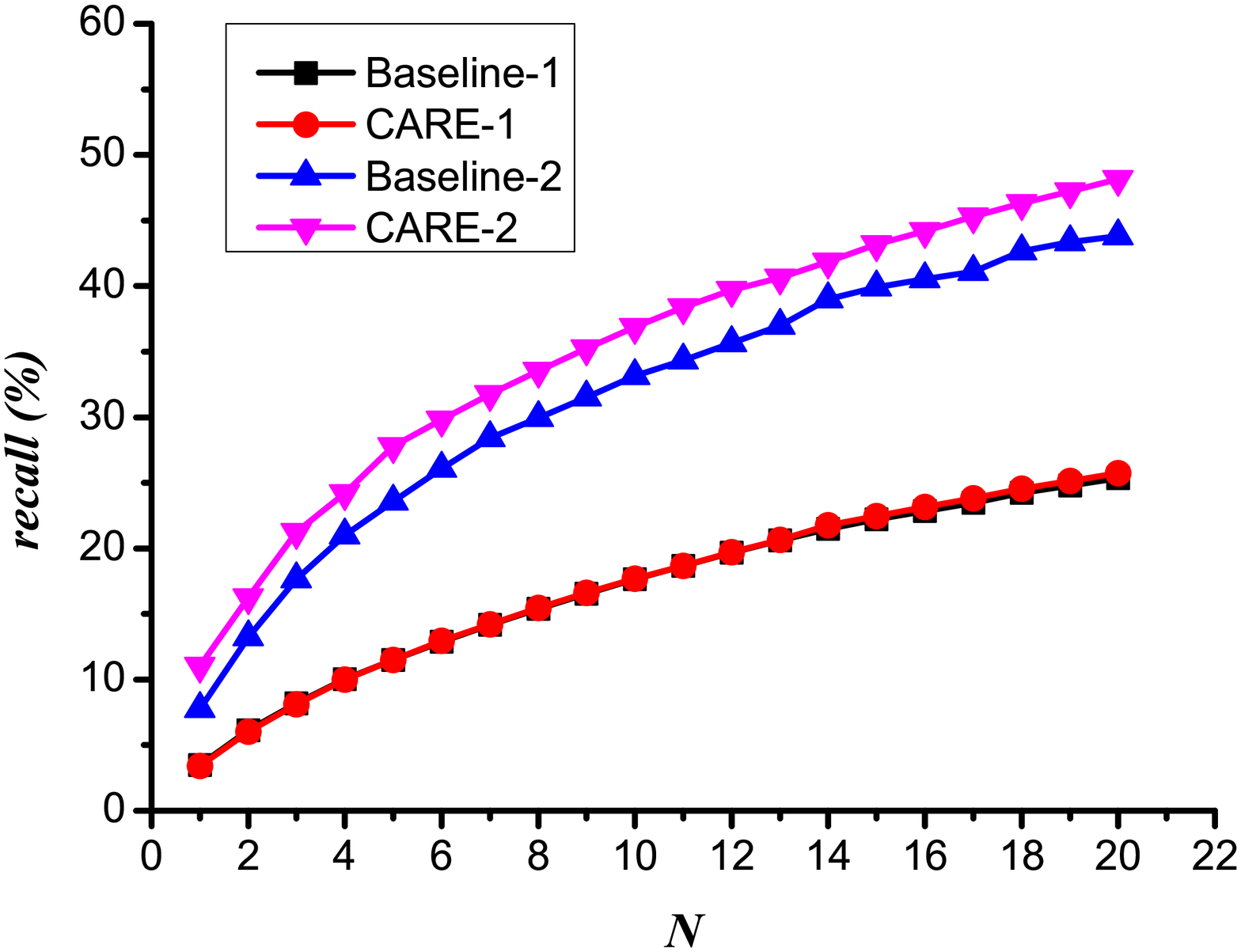}}
\subfigure[F1]{
\label{f1}
\includegraphics[width=0.325\textwidth]{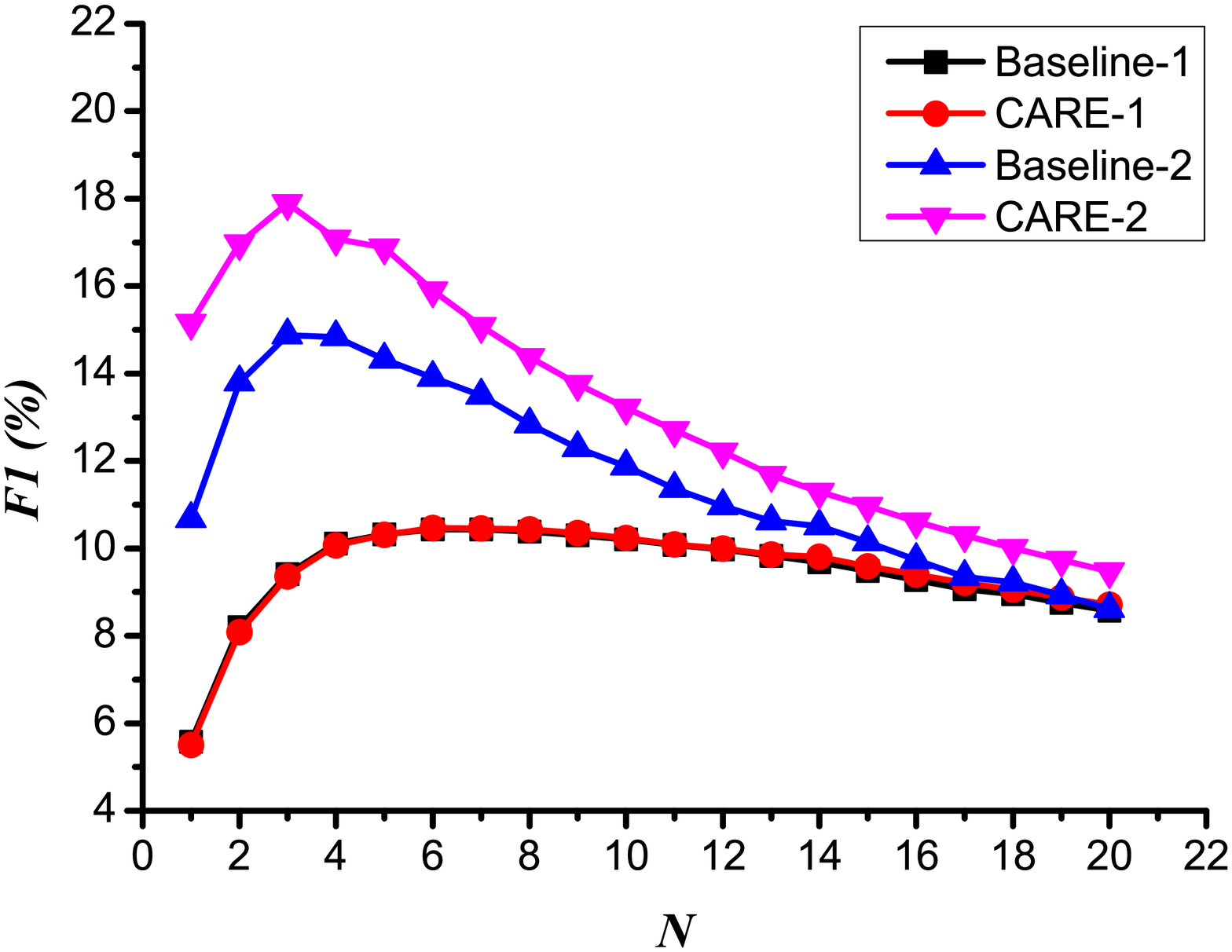}}
\caption{Comparison of precision, recall, and F1 of Baseline and CARE for all researchers and relevant researchers.}
\label{comparison}
\end{figure*}
\subsection{Experimental Setup}
To test our method's performance, the dataset is randomly divided into a training set (80\%) and a test set (20\%) using the following procedures. For researchers who have less than 5 articles in his library, we randomly select one article and the corresponding researcher into the test set. For others, we randomly select articles into the test set at the ratio of 20\%. The training set is treated as known information used by our method for generating recommendations, while the test set is regarded as unknown information used for testing the performance of recommendation results. To evaluate the recommendation quality of our proposed method, in our experiments, we employed three different metrics, namely, Precision, Recall, and F1, which have been widely used in the literatures \cite{herlocker2004evaluating,bobadilla2013recommender,grossman2012information} on the fields of recommender systems and information retrieval. Next, we give their definition information.

$\bullet$
\textit{Precision}. Precision represents the probability that the selected articles appeared in the recommendation list which is shown as
\begin{equation}
\begin{aligned}
Precision_i=\frac{N_{rt}^i}{N}
\end{aligned}
\label{precision-i}
\end{equation}
where $Precision_i$ represents researcher $R_i$'s precision, $N_{rt}^i$ denotes the number of recommended articles that appeared in the researcher $R_i$'s test set, and $N$ represents the length of recommendation list. By averaging over all researchers' precisions, we can obtain the whole recommender systems' precision as
\begin{equation}
\begin{aligned}
Precision=\frac{1}{n}\sum_{i=1}^n Precision_i
\end{aligned}
\label{precision}
\end{equation}
where $n$ represents the number of researchers. Obviously, a higher precision means a higher recommendation accuracy.

$\bullet$
\textit{Recall}. Recall represents the probability that the recommended articles appeared in researcher's collected list shown as
\begin{equation}
\begin{aligned}
Recall_i=\frac{N_{rt}^i}{N_t^i}
\end{aligned}
\label{recall-i}
\end{equation}
where $Recall_i$ represents researcher $R_i$'s recall and $N_t^i$ is the number of articles collected by researcher $R_i$ in the test set. Averaging over all individuals' recall, we can obtain the whole recommender systems' recall as
\begin{equation}
\begin{aligned}
Recall=\frac{1}{n}\sum_{i=1}^n Recall_i
\end{aligned}
\label{precision}
\end{equation}

$\bullet$
\textit{F1}. Generally speaking, for each researcher, recall is sensitive to $N$ and a larger value of $N$ generally gives a higher recall but a lower precision. F1, that assigns equal weight for precision and recall, is defined as
\begin{equation}
\begin{aligned}
F1_i=\frac{2 \times Precision_i \times Recall_i}{Precision_i+Recall_i}
\end{aligned}
\label{F1-i}
\end{equation}
By averaging over all researchers' F1, we can also obtain the whole system's F1 as
\begin{equation}
\begin{aligned}
F1=\frac{1}{n}\sum_{i=1}^n F1_i
\end{aligned}
\label{F1}
\end{equation}

In order to demonstrate the effectiveness of our recommendation method, we compare CARE with the following method.

$\bullet$
\textbf{Baseline}: This is a random walk model with restart, which does not take into account common author relations between articles and does not differentiate between relevant researchers and irrelevant researchers.

\begin{figure*}
\centering
\subfigure[Precision]{
\label{precision}
\includegraphics[width=0.325\textwidth]{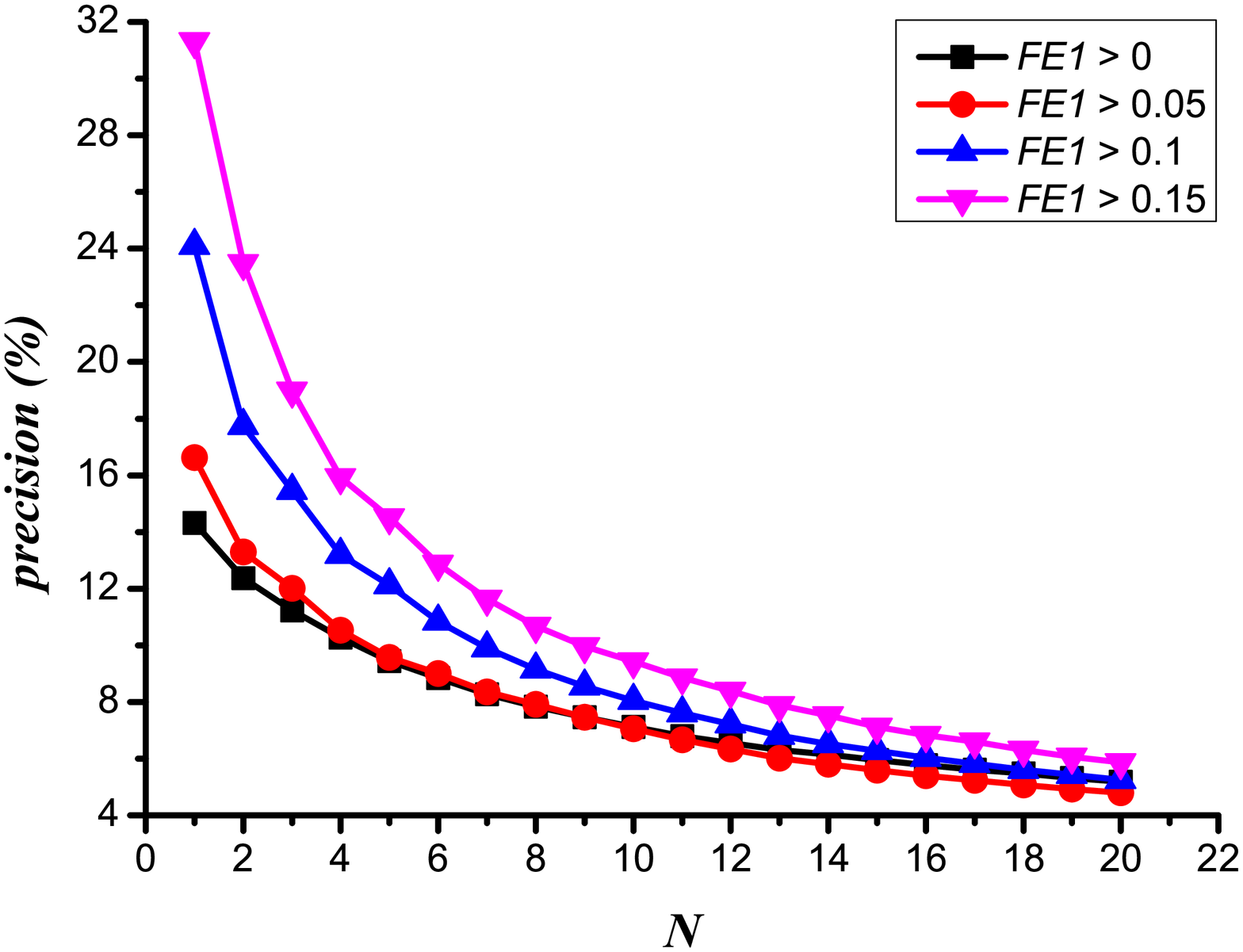}}
\subfigure[Recall]{
\label{recall}
\includegraphics[width=0.325\textwidth]{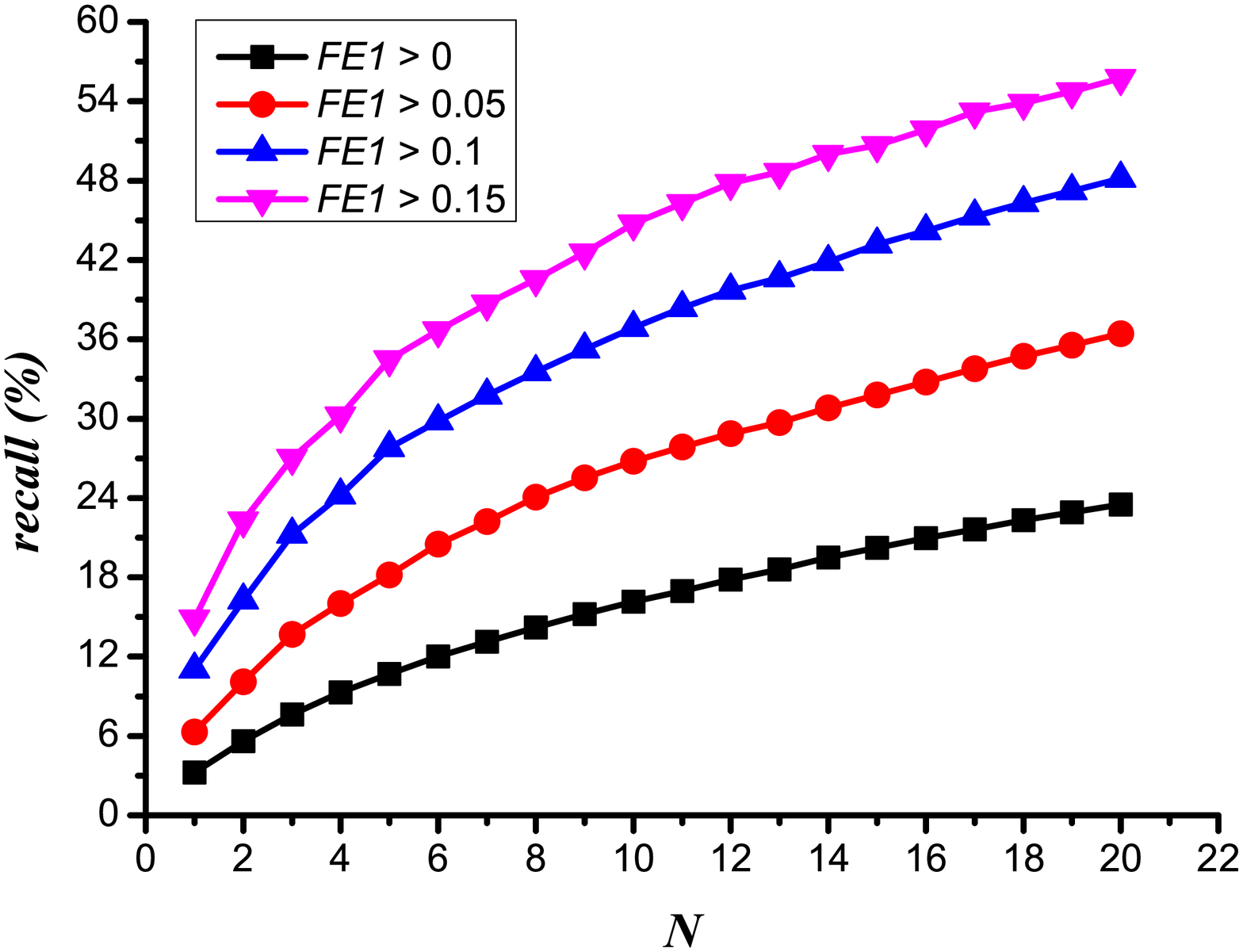}}
\subfigure[F1]{
\label{f1}
\includegraphics[width=0.325\textwidth]{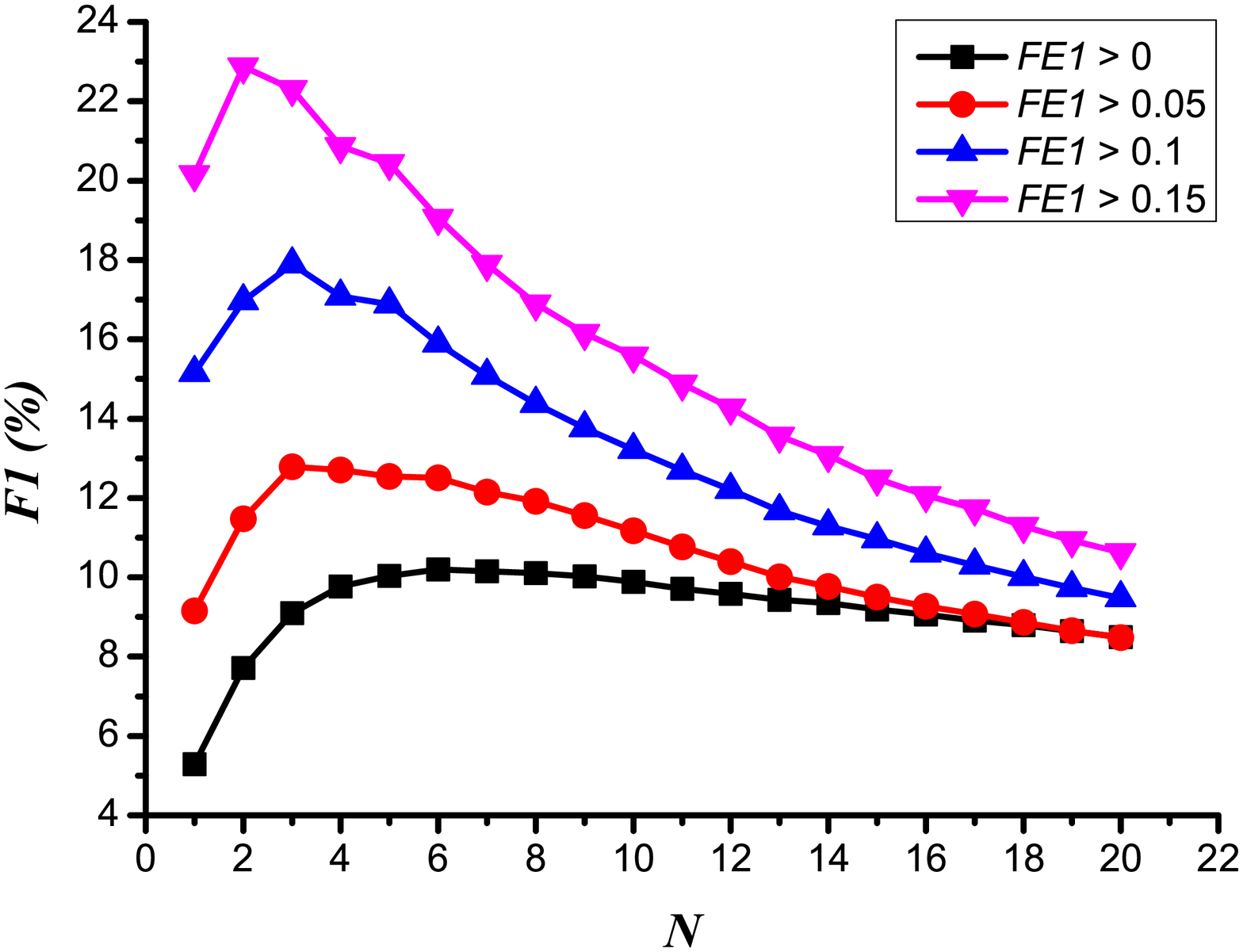}}
\caption{Comparison of precision, recall, and F1 of CARE for different thresholds of FE1.}
\label{FE1}
\end{figure*}

\begin{figure*}
\centering
\subfigure[Precision]{
\label{precision}
\includegraphics[width=0.325\textwidth]{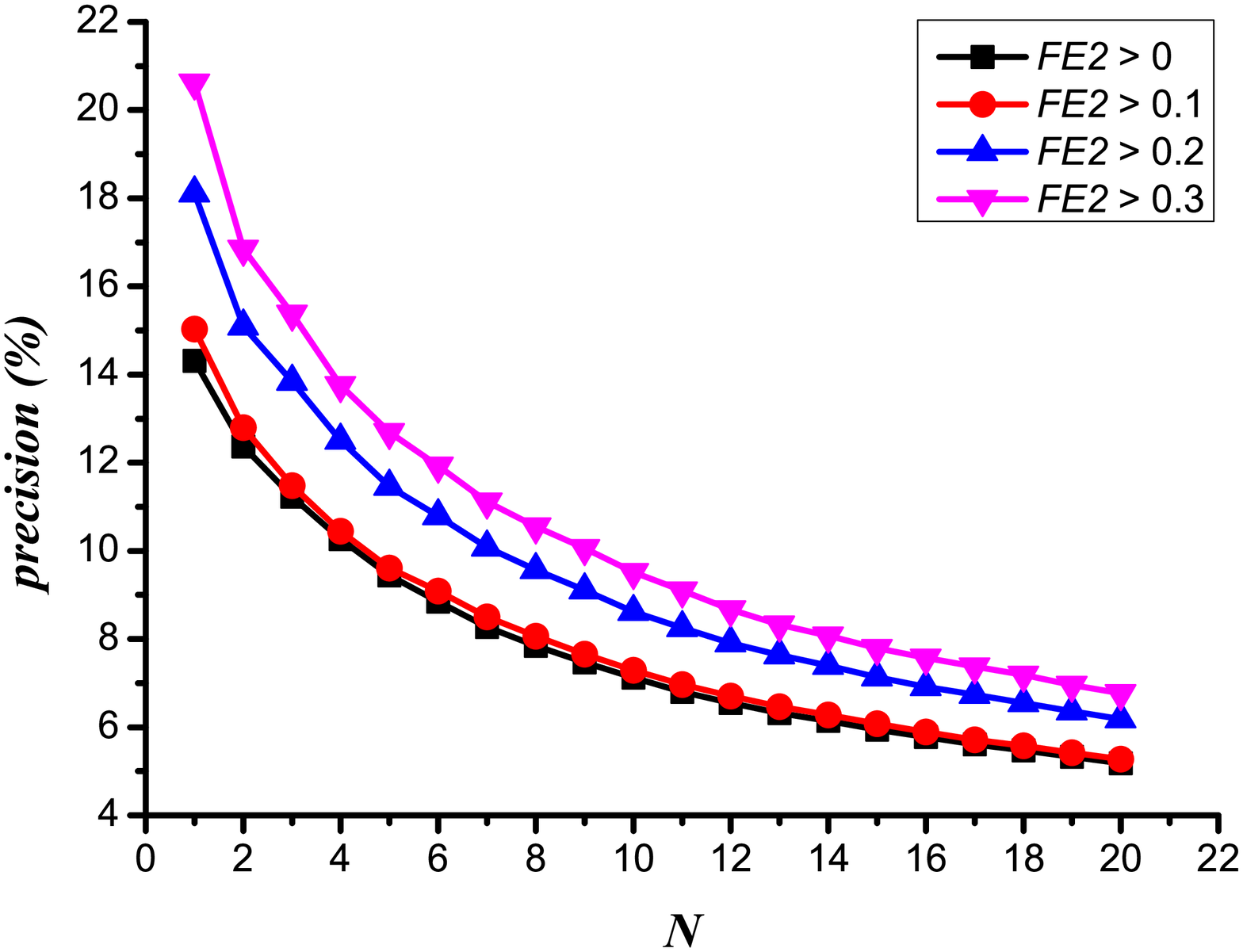}}
\subfigure[Recall]{
\label{recall}
\includegraphics[width=0.325\textwidth]{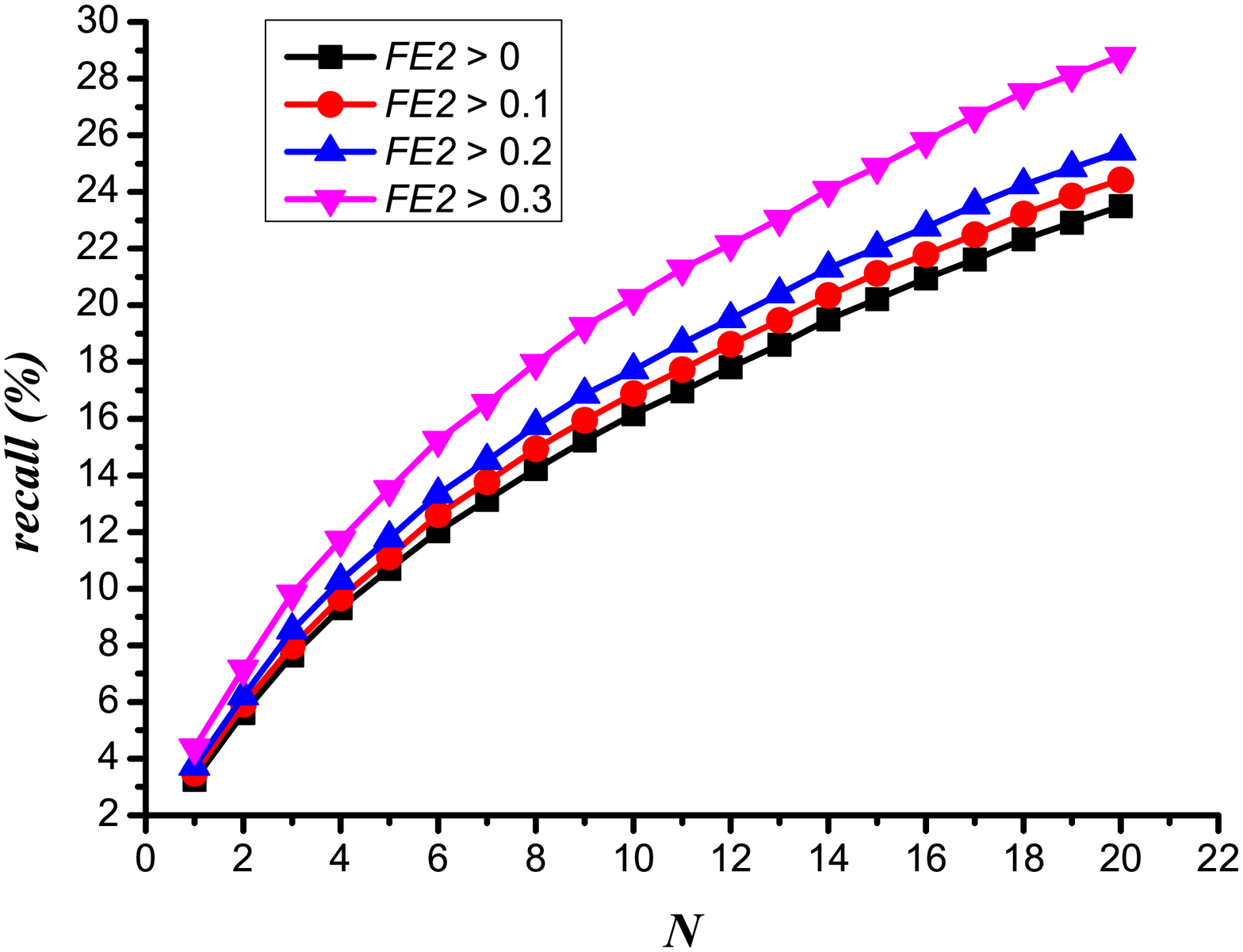}}
\subfigure[F1]{
\label{f1}
\includegraphics[width=0.325\textwidth]{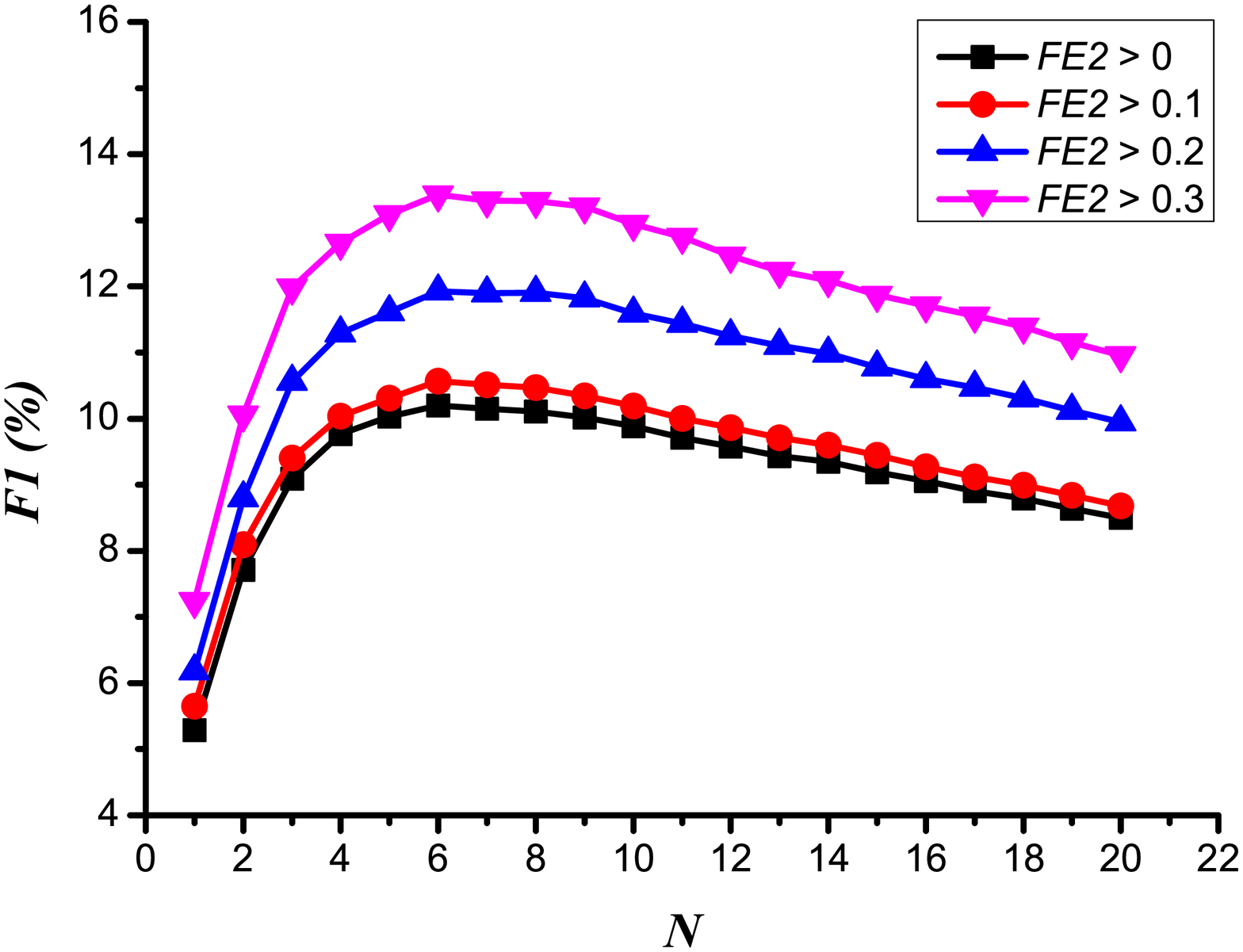}}
\caption{Comparison of precision, recall, and F1 of CARE for different thresholds of FE2.}
\label{FE2}
\end{figure*}

%

\begin{figure*}
\begin{center}
\begin{minipage}[c]{0.5\textwidth}
\centering
\includegraphics[angle=0,width=9cm]{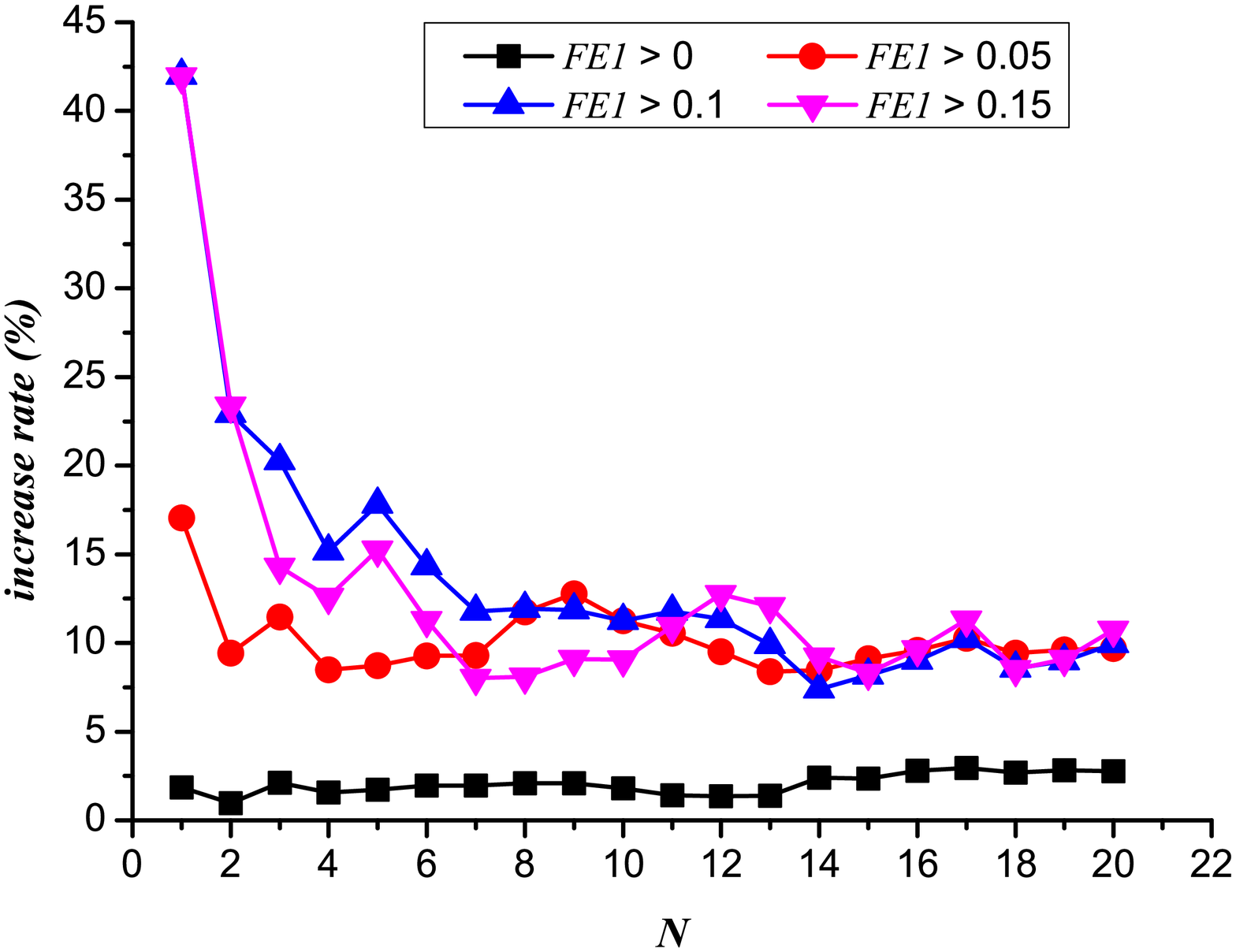}
\centering
\caption{Impact of FE1 on increase rate.}
\label{increase-FE1}
\end{minipage}%
\begin{minipage}[c]{0.5\textwidth}
\centering
\includegraphics[angle=0,width=9cm]{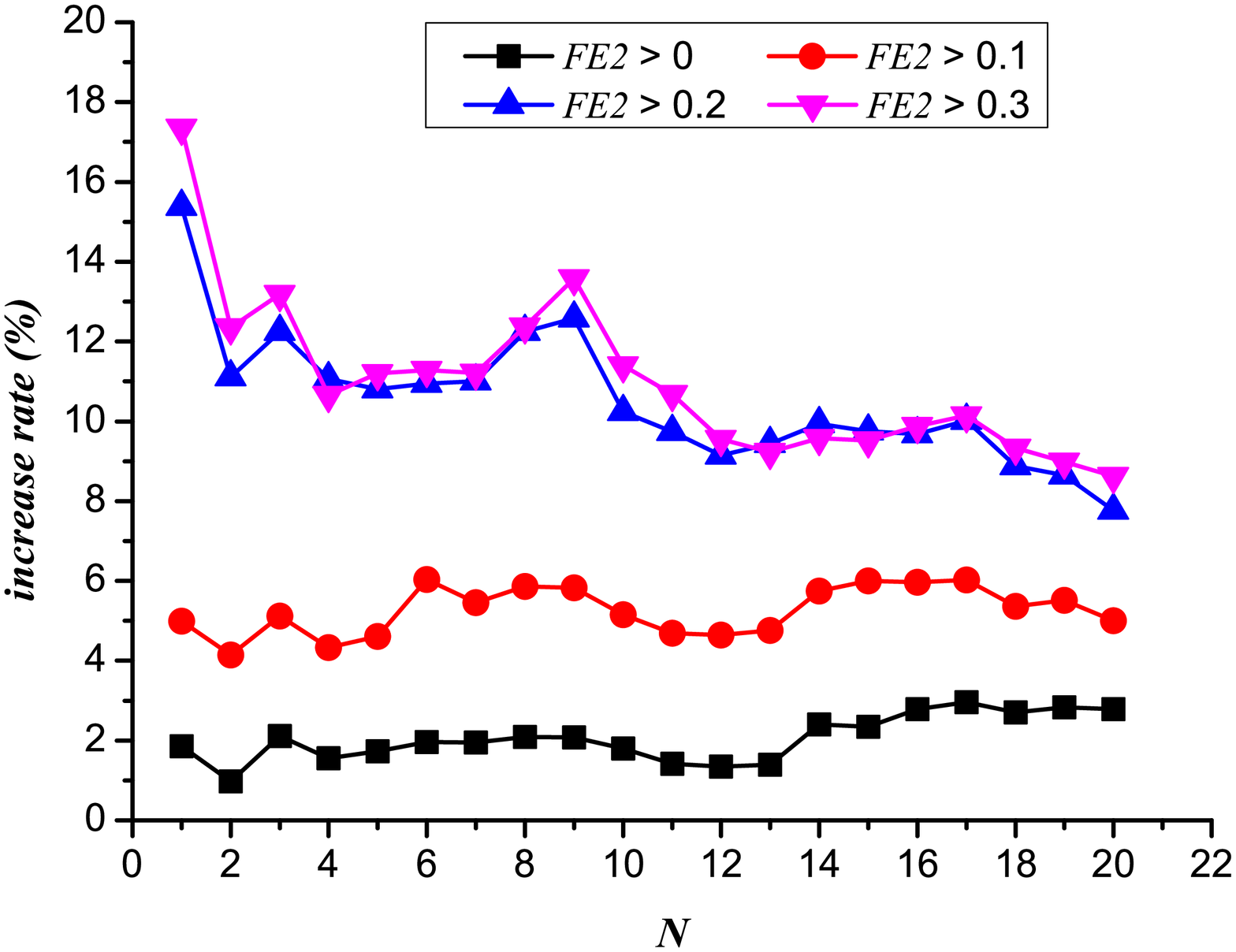}
\centering
\caption{Impact of FE2 on increase rate.}
\label{increase-FE2}
\end{minipage}
\end{center}
\end{figure*}

\subsection{Impact of Walking Probability}
As stated in Section 4.3, for a ceratin vertex, $\alpha$ represents the walking probability from the vertex to its neighbor vertices and ($1-\alpha$) represents the walking probability from the vertex to the source vertex (target researcher). Different values of $\alpha$ may produce different impacts on recommendation quality. We conducted relevant experiments using our proposed CARE method for different values of $\alpha$. Fig. \ref{alpha-impact} shows the comparison results of precision, recall, and F1 when $\alpha$ is equal to 0.2, 0.4, 0.6, and 0.8, respectively. As shown in these sub-figures, for a larger value of $\alpha$, our CARE method achieves larger values of precision, recall, and F1. For example, when N is equal to 6, CARE ($\alpha$ is equal to 0.2) achieves the worst results (6\% precision, 10\% recall, and 7.5\% F1), and CARE ($\alpha$ is equal to 0.8) achieves the best results (8.5\% precision, 14\% recall, and 11\% F1). This indicates that different walking probabilities have different impacts on CARE method. However, because $\alpha$ is the common parameter for CARE and Baseline methods, it is enough to discuss their comparison results only if they employ the same value of $\alpha$. Therefore, we assign an empirical value of 0.8 to $\alpha$ for next experiments.

\subsection{Comparison against Baseline Method}
\begin{figure*}
\centering
\subfigure[Precision]{
\label{precision}
\includegraphics[width=0.325\textwidth]{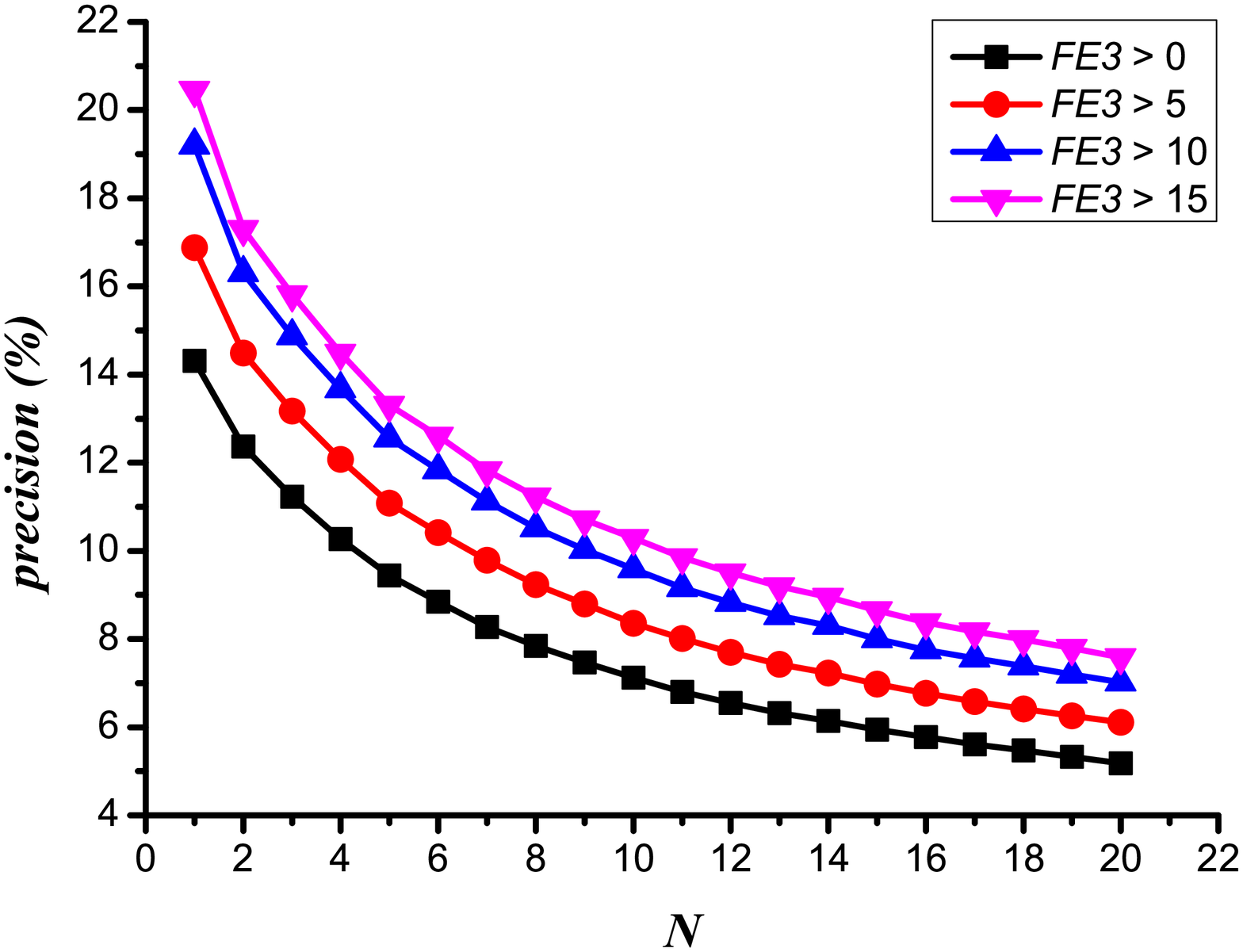}}
\subfigure[Recall]{
\label{recall}
\includegraphics[width=0.325\textwidth]{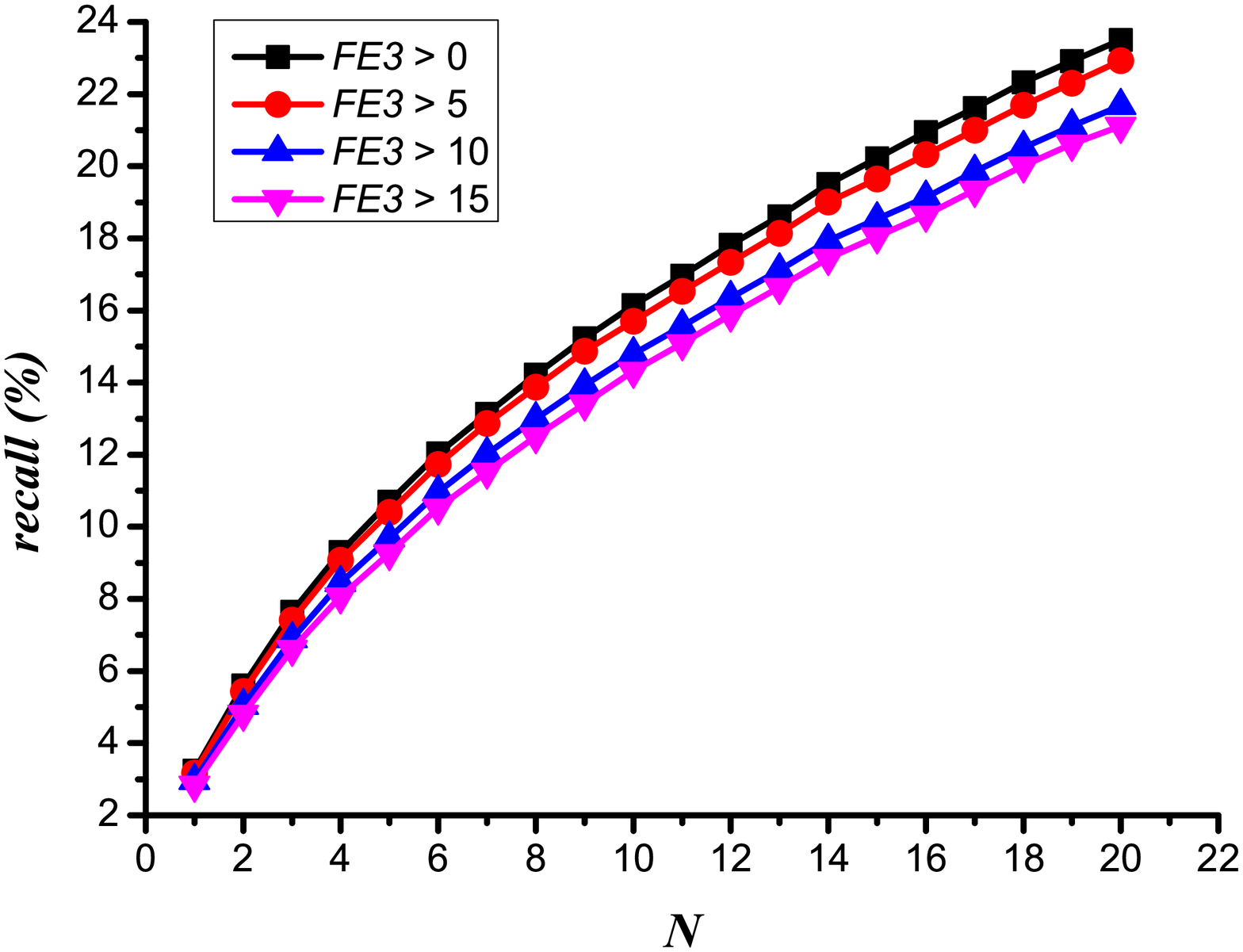}}
\subfigure[F1]{
\label{f1}
\includegraphics[width=0.325\textwidth]{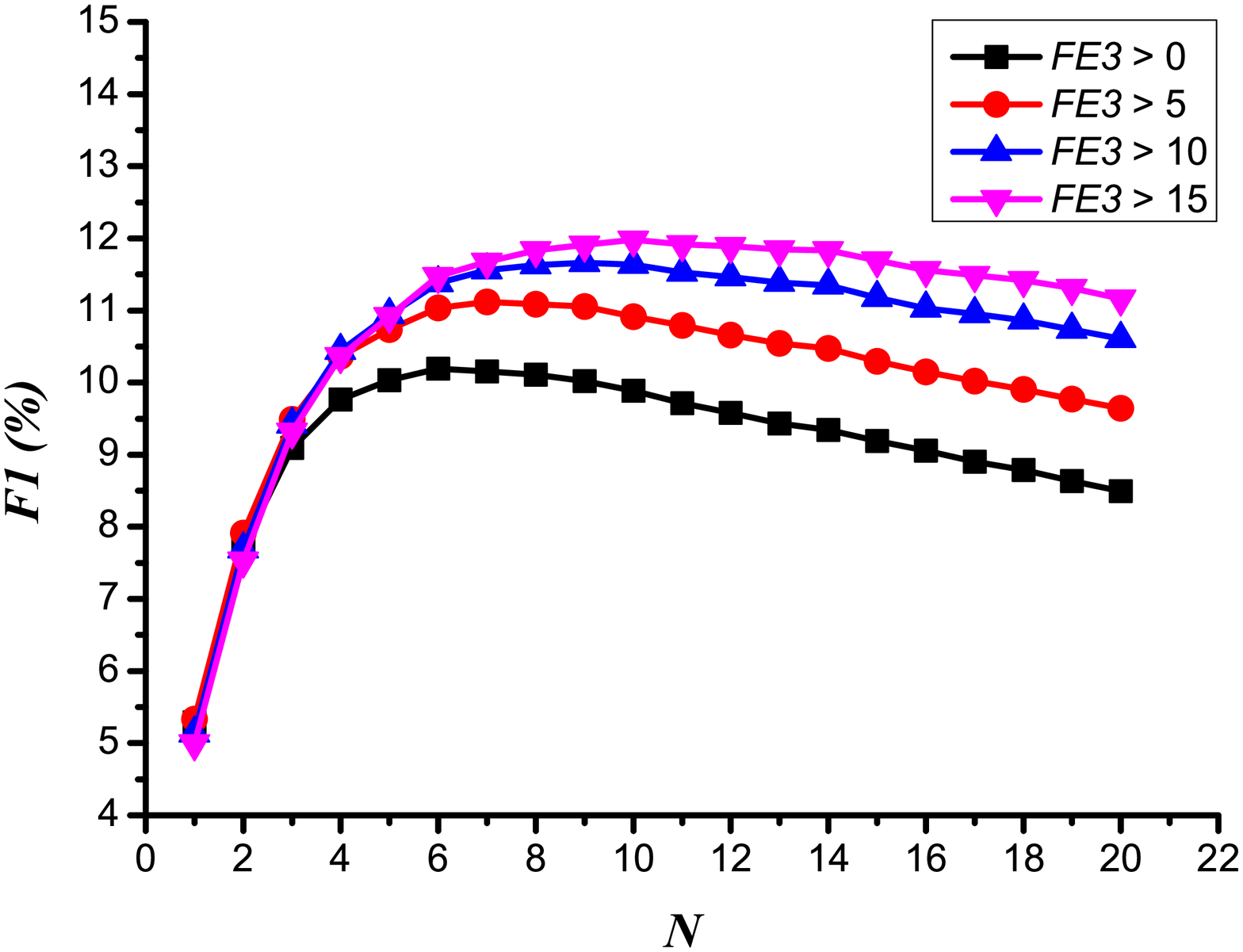}}
\caption{Comparison of precision, recall, and F1 of CARE for different thresholds of FE3.}
\label{FE3}
\end{figure*}

\begin{figure*}
\centering
\subfigure[Precision]{
\label{precision}
\includegraphics[width=0.325\textwidth]{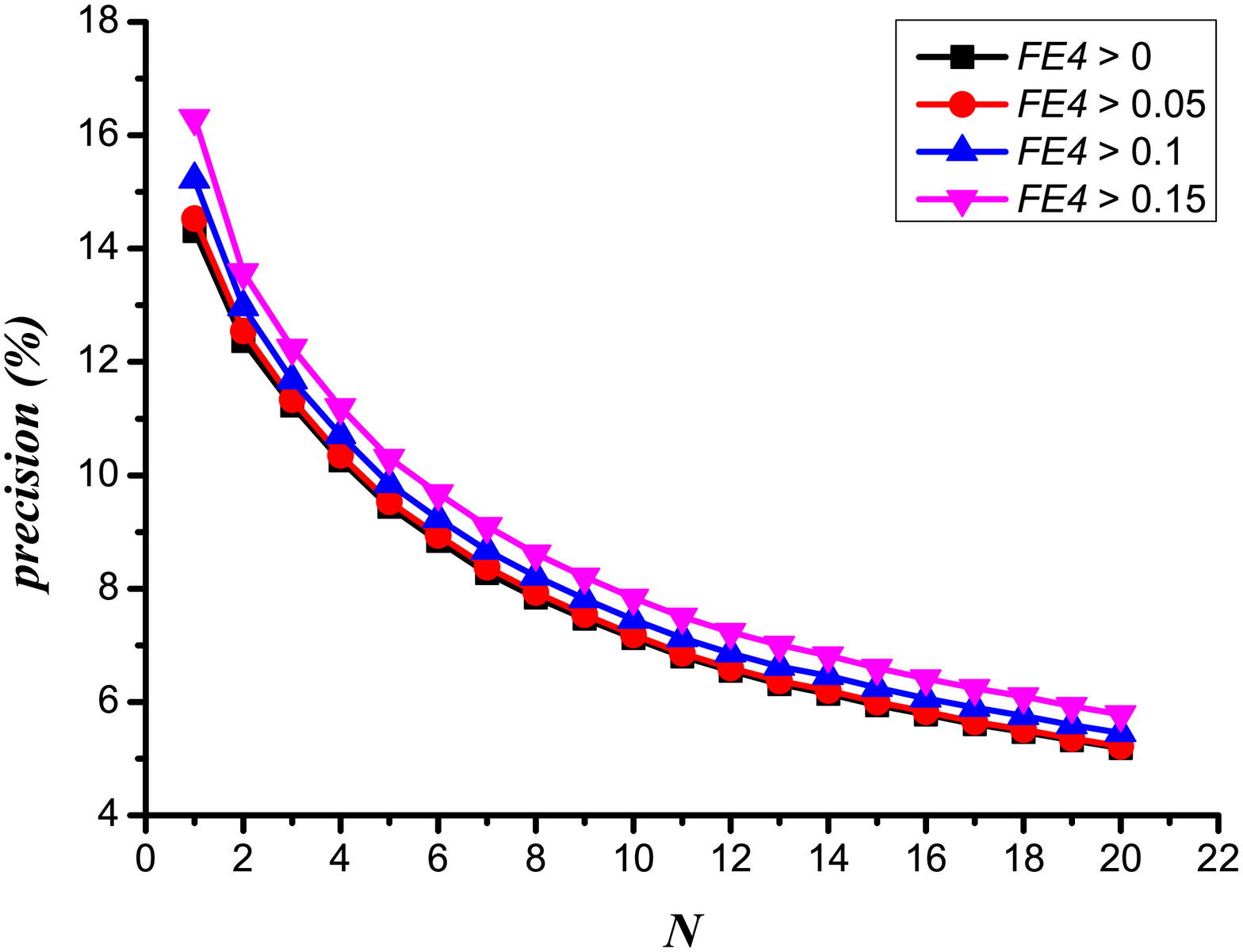}}
\subfigure[Recall]{
\label{recall}
\includegraphics[width=0.325\textwidth]{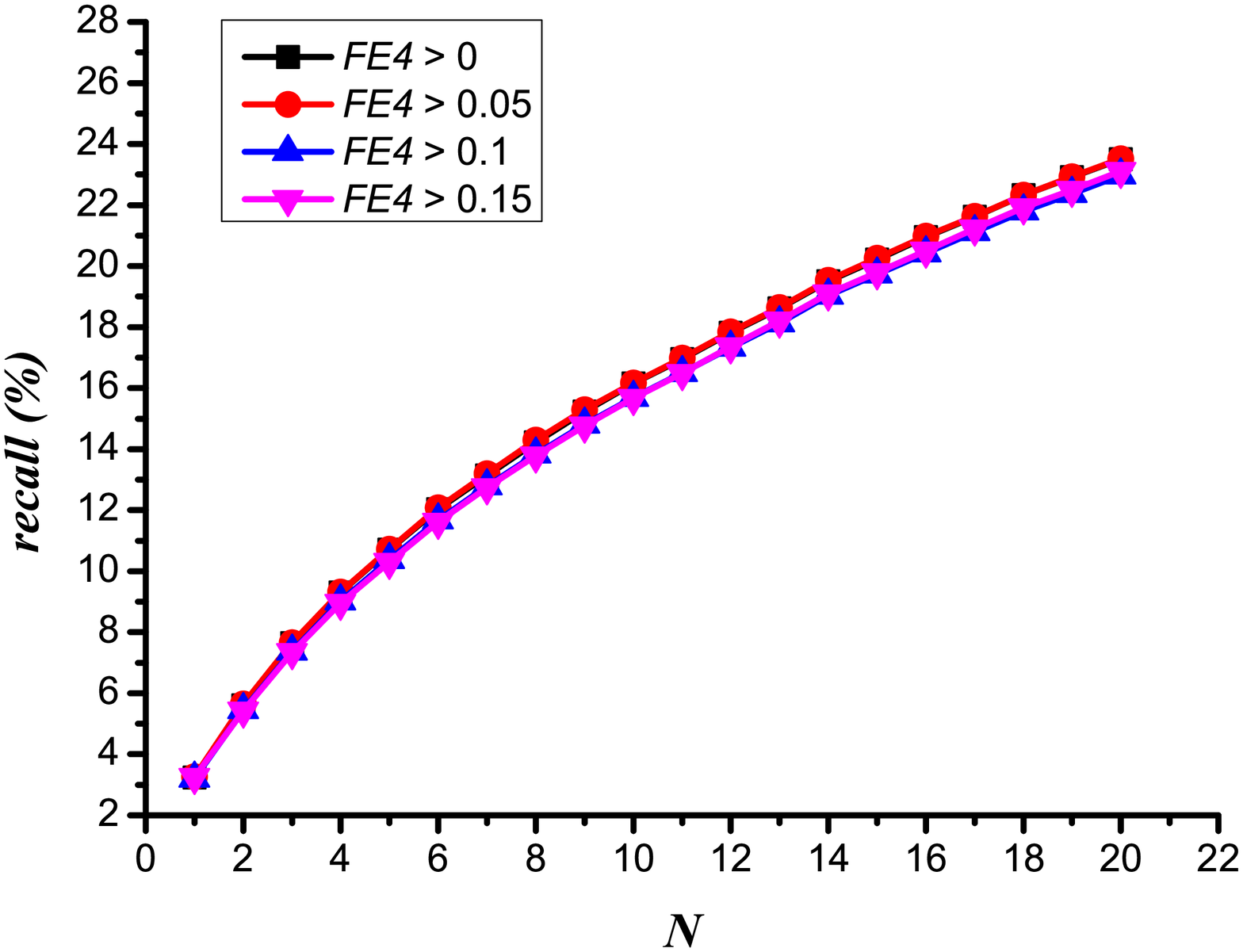}}
\subfigure[F1]{
\label{f1}
\includegraphics[width=0.325\textwidth]{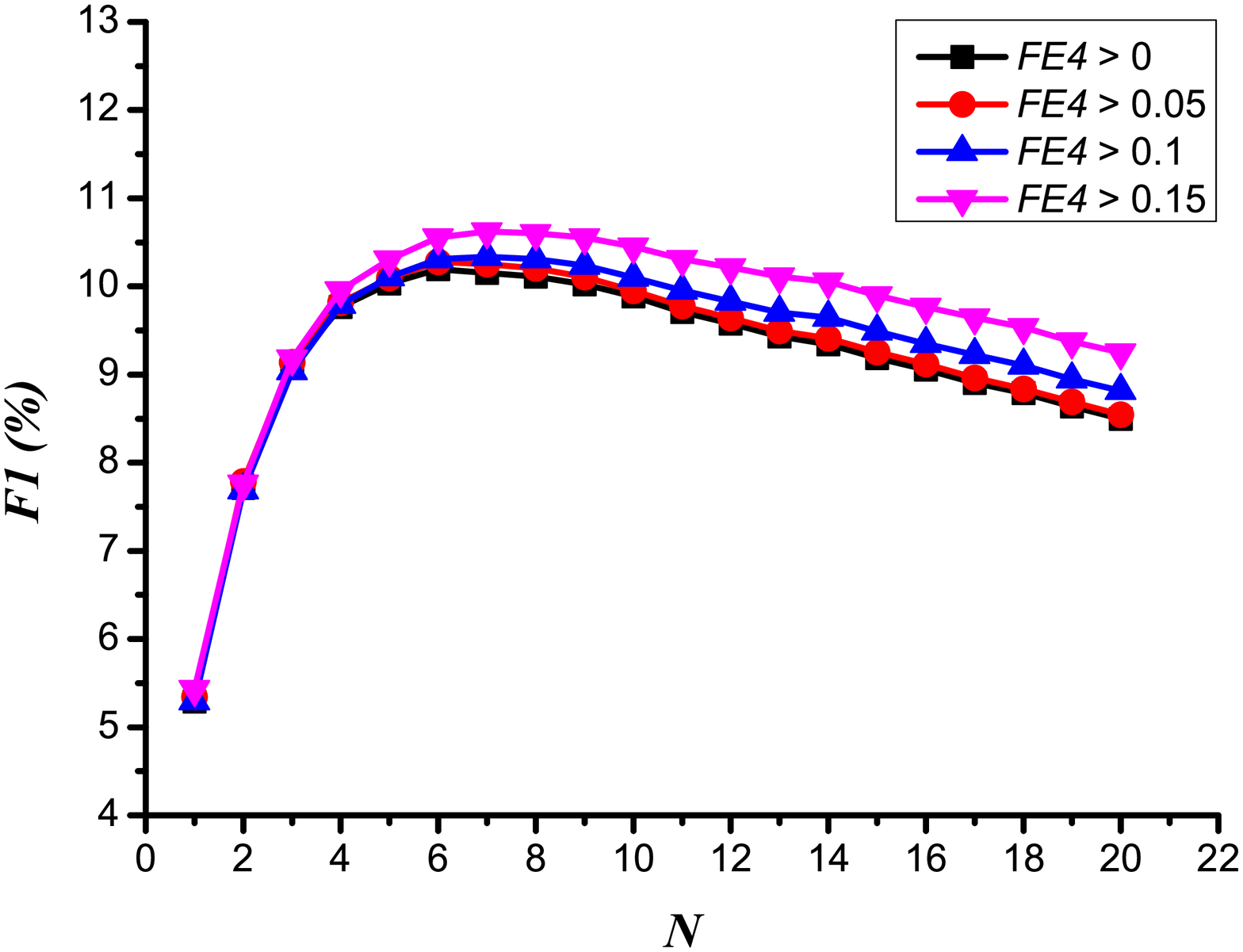}}
\caption{Comparison of precision, recall, and F1 of CARE for different thresholds of FE4.}
\label{FE4}
\end{figure*}

In this section, we conducted two groups of experiments: (i) compare the results of CARE and Baseline when all researchers are targets and the two methods in this situation are called CARE-1 and Baseline-1; (ii) compare the results of CARE and Baseline when a part of researchers are selected as relevant targets through the previously-defined features ($FE1>0.1$ and $FE2>0.1$) and the two methods under this situation are called CARE-2 and Baseline-2. Fig. \ref{comparison} shows the result comparisons of the two groups of experiments. In the comparison of CARE-1 to Baseline-1, it can be obviously seen that, their precision, recall, and F1 are almost the same. Actually, the results of CARE-1 are a little worse than those of Baseline-1. However, we can see that, precision, recall, and F1 of CARE-2 method are significantly larger than those of Baseline-2 method. It means that, CARE performs better than Baseline method for relevant researchers filtered using two features $FE1$ and $FE2$. This indicates that, incorporating common author relations is able to help generate accurate recommendations for relevant researchers rather than all researchers.

\subsection{Impact of Researcher Features}
In this section, we conducted relevant experiments to discuss the impact of researcher's features previously defined in Section 4.2 on recommendation quality of CARE method. Fig. \ref{FE1} shows the comparison results of CARE for different thresholds of $FE1$ (i.e., 0, 0.05, 0.1, and 0.15). We can see that, for a larger threshold of $FE1$, CARE achieves larger precision, recall, and F1. For example, for the length of recommendation list is equal to 2, when the threshold of $FE1$ is equal to 0, their results are 12\% precision, 6\% recall, and 7\% F1 (i.e., the least values), and when the threshold of $FE1$ is equal to 0.15, their results are 23\% precision, 22\% recall, and 23\% F1 (i.e., the largest values). This indicates that, a larger threshold of $FE1$ can help to find more relevant researchers with author-based search patterns and then our proposed CARE method generates better recommendations for these targets. In addition, Fig. \ref{FE2} shows the comparison results of CARE for different thresholds of $FE2$ (i.e., 0, 0.1, 0.2, and 0.3). We can also see that, CARE with a larger threshold of $FE2$ performs better than that with a smallest threshold of $FE2$. Similarly, this demonstrates that, $FE2$ is an effective feature for determining relevant target researchers who have author-based searcher patterns.

In addition, we defined \textit{increase rate} to represent CARE's improvement ratio to Baseline for different thresholds of $FE1$ and $FE2$ using Equation (\ref{increase-rate}). Note that \textit{increase rate} is the same for precision, recall, and F1. Fig. \ref{increase-FE1} shows the \textit{increase rate} when the thresholds of $FE1$ are 0, 0.05, 0.1, and 0.15, respectively. It can be observed that, the \textit{increase rate} is positive for these four situations. This demonstrates that, for the same researchers who are filtered using the threshold of $FE1$, CARE performs better than the Baseline method. Fig. \ref{increase-FE2} shows the \textit{increase rate} when the thresholds of $FE2$ are 0.1, 0.2, 0.3, and 0.4. We can also see that, the \textit{increase rate} is positive for all situations. Especially, when the threshold of $FE2$ is larger (e.g., 0.3 or 0.4), CARE performs much better than the Baseline method. These experiments further illustrate that these two features are useful to help find researchers with author-based search patterns and our CARE method is effective in terms of generating better recommendations for those targets.
\begin{equation}
\begin{aligned}
increase\text{ }rate
&=\frac{F1(CARE) - F1(Baseline)}{F1(Baseline)}
\end{aligned}
\label{increase-rate}
\end{equation}

What's more, we also defined the following two features.

$\bullet$
$FE3$, is the total number of pairwise articles with common author relations for a researcher.

$\bullet$
$FE4$, is the ratio of the number of common authors who exist in all articles to the total number of articles for a researcher.

We also use Fig. \ref{ComputationExample} as an example scenario for illustrating the computation process of $FE3$ and $FE4$. From Fig. \ref{ComputationExample}(b), we can easily obtain the number of pairwise articles, therefore $FE3$ is equal to 4. Additionally, in the example scenario, the common authors are $U1$, $U2$, $U4$, and $U5$. Then, $FE4$ is equal to $4/4=1$. Using $FE3$ and $FE4$, we conducted relevant experiments as shown in Figs. \ref{FE3} and \ref{FE4}. For a larger value of $FE3$ or $FE4$, F1 of CARE also achieves a larger value, but recall is smaller in Fig. \ref{FE3}(b) or almost the same in Fig. \ref{FE4}(b). There may be the following reasons: (i) The feature $FE3$ takes into account the number of existing common author relations between articles but ignores the total number of articles read by a researcher; (ii) The feature $FE4$ does not take into account the occurrence number of each common author. Therefore, the two features are considered to be ineffective for determining relevant researchers.

\section{Conclusion}
In this paper, a novel method that exploits information pertaining to common author relations and historical preferences has been proposed to recommend articles of interest for specific researchers with author-based search patterns. In order to determine specific targets, we defined two features (i.e. $FE1$ and $FE2$) which are relevant to common author relations between articles. Then, the information on common authors relations was incorporated to build a graph-based article ranking algorithm for generating a recommendation list. The experimental results demonstrated that, for relevant targets determined by two features, our proposed method performs better than the Baseline method and the two features have impacts on recommendation quality. In addition, we also defined two other features ($FE3$ and $FE4$ in Section 5.5) and they are proved to be ineffective for suitable targets selection through relevant experiments.

As our future work, we plan to define new features to explore which researchers have author-based search patterns. In addition, it is potential to incorporate additional social relations such as citation relationships to design a citation-based recommendation method. Then, relevant targets are determined by analyzing the information on citation relations between articles. Finally, different recommendation methods which are suitable for different researchers can be combined into a hybrid framework so that all researchers can obtain satisfactory recommendations.

\ifCLASSOPTIONcaptionsoff
  \newpage
\fi

\bibliographystyle{IEEEtran}
\bibliography{IEEEabrv}

\begin{IEEEbiography}[{\includegraphics[width=1in,height=1.25in,clip,keepaspectratio]{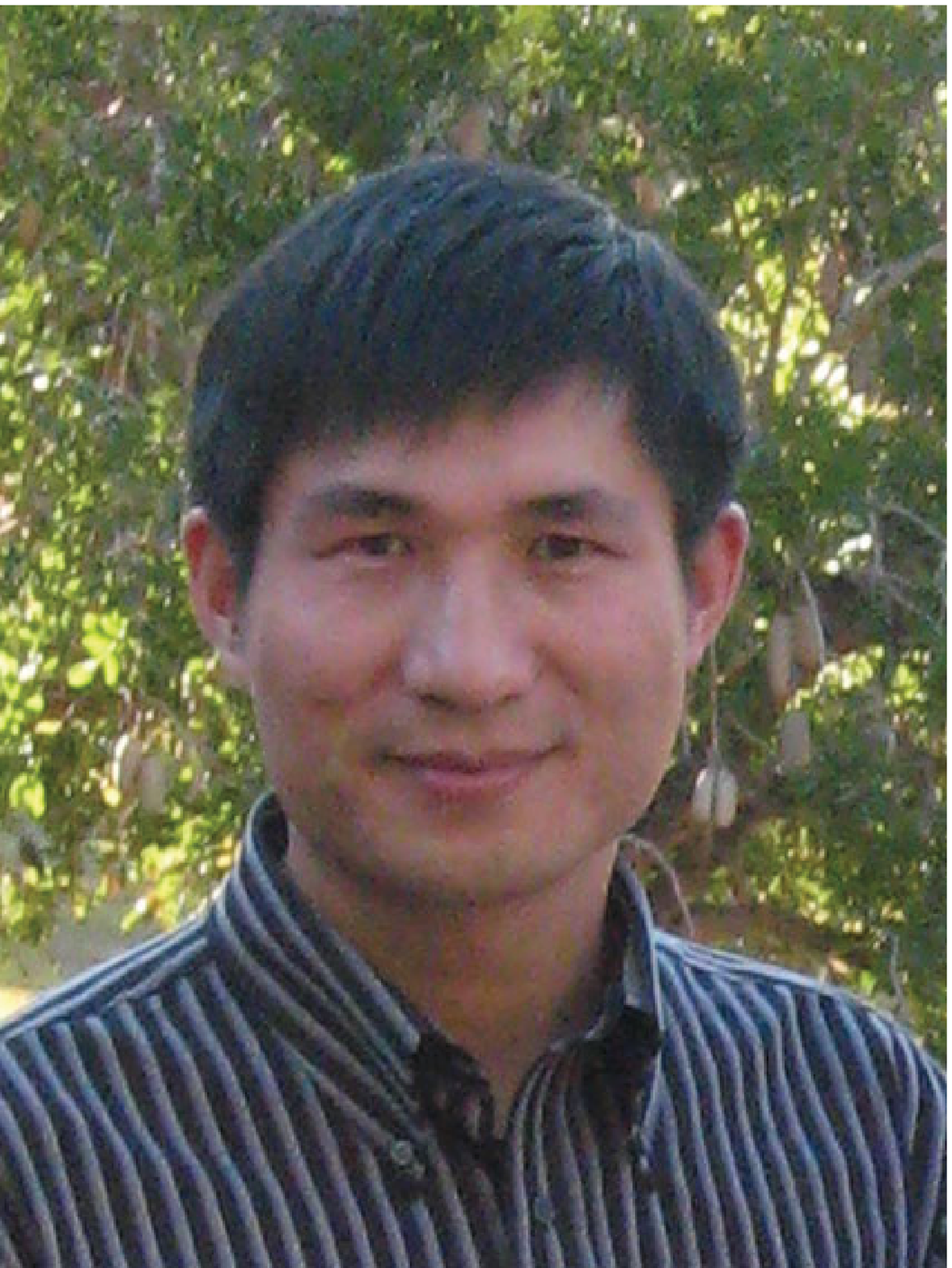}}]{Feng Xia (M'07-SM'12)}
received the BSc and Ph.D. degrees from Zhejiang University, Hangzhou, China. He was a Research Fellow at Queensland University of Technology, Australia. He is currently a Full Professor in School of Software, Dalian University of Technology, China. He is the (Guest) Editor of several international journals. He serves as General Chair, PC Chair, Workshop Chair, Publicity Chair, or PC Member of a number of conferences. Dr. Xia has authored/co-authored two books and over 200 scientific papers in international journals and conferences. His research interests include computational social science, big data, and mobile social networks. He is a Senior Member of IEEE (Computer Society, SMC Society) and ACM (SIGWEB), and a Member of AAAS.
\end{IEEEbiography}
\begin{IEEEbiography}[{\includegraphics[width=1in,height=1.25in,clip,keepaspectratio]{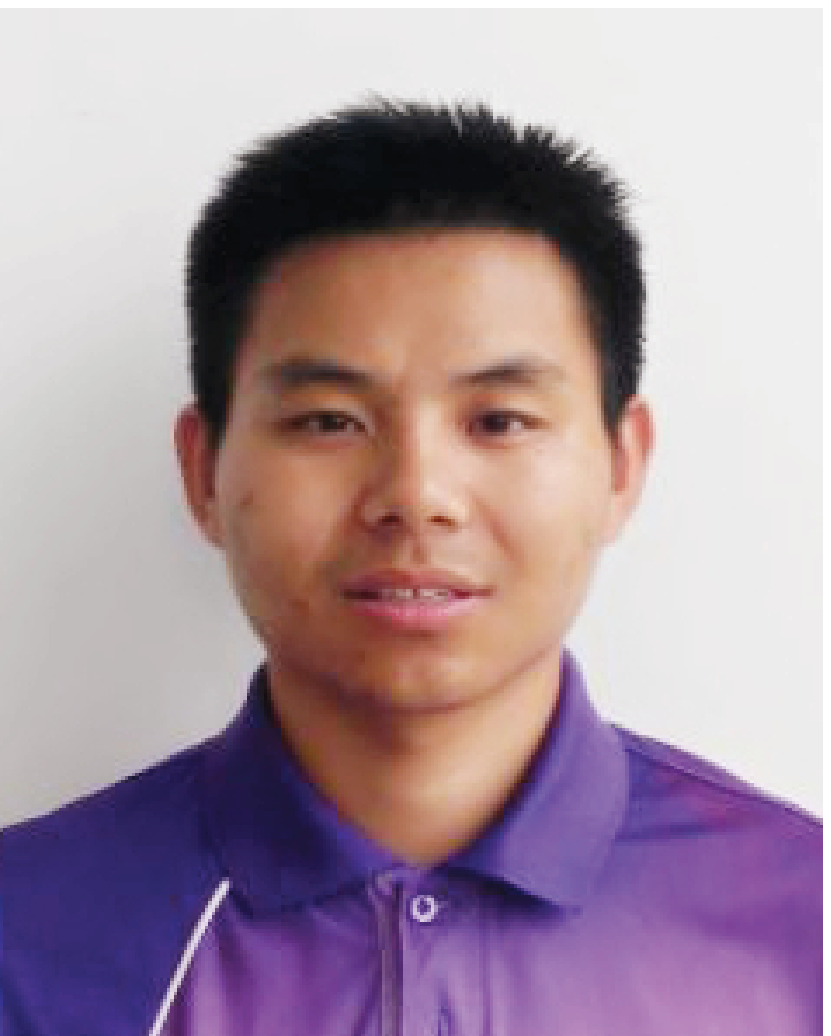}}]{Haifeng Liu}
 received his BSc and MSc degrees in Computer Science from Dalian University of Technology, China, in 2009 and 2012, respectively. He is currently pursuing his PhD in School of Software, Dalian University of Technology. His research interests include recommender systems, mobile computing and big data.
\end{IEEEbiography}
\begin{IEEEbiography}[{\includegraphics[width=1in,height=1.25in,clip,keepaspectratio]{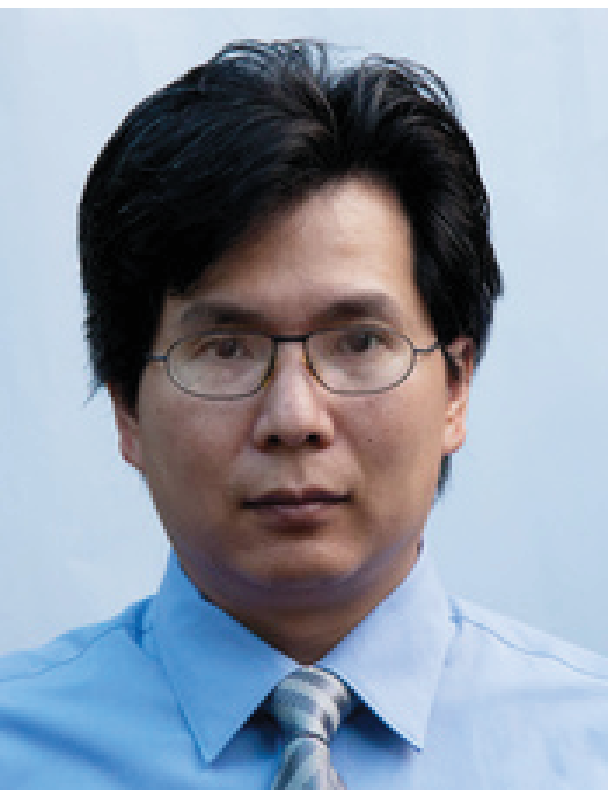}}]{Ivan Lee (SM'07)}
 received Bachelor of Engineering (Hons), Master of Commerce, Master of Engineering by Research, and PhD degrees from the University of Sydney, Australia. Between 1998 and 2001, he worked for Cisco Systems as a software development engineer. Between 2003 and 2005, he worked for Remotek Corporation as a software engineer at the advanced technology group. Between 2005 and 2007, he was an Assistant Professor in the Department of Electrical and Computer Engineering, Ryerson University, Canada. Since 2008, he has been a Senior Lecturer in the School of IT and Mathematical Sciences at the University of South Australia. His research interests include multimedia systems, medical imaging, data analytics, and computational economics.
\end{IEEEbiography}
\begin{IEEEbiography}[{\includegraphics[width=1in,height=1.25in,clip,keepaspectratio]{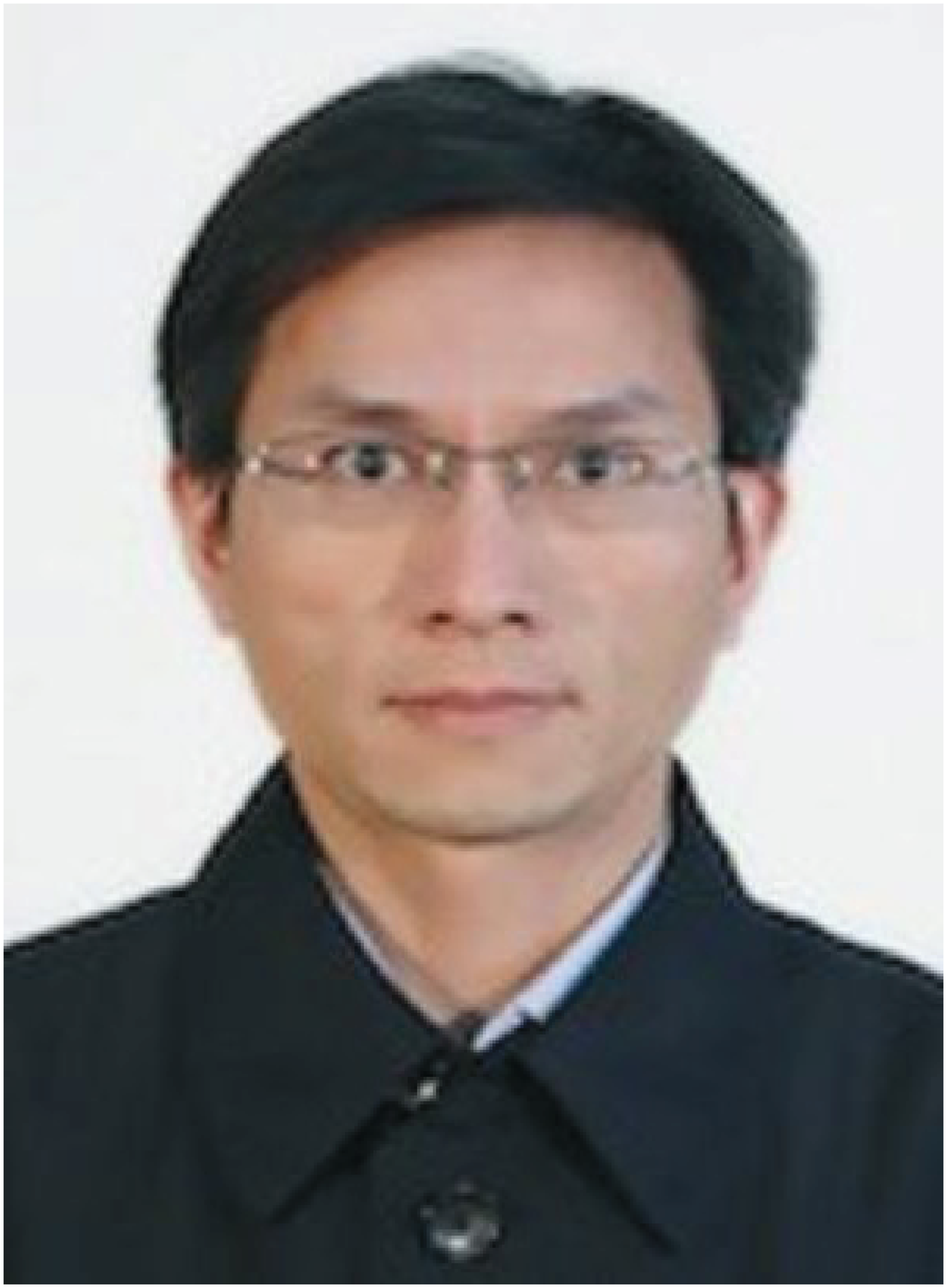}}]{Longbing Cao (SM'06)}
 received the Ph.D. degree in Pattern Recognition and Intelligent Systems in Chinese Academy of Sciences and another in Computing Science from the University of Technology Sydney (UTS). He is currently a Professor of information technology with the University of Technology Sydney (UTS), Sydney, Australia, where he is the Founding Director of the Advanced Analytics Institute. He is also the Research Leader of the Data Mining Program with the Australian Capital Markets Cooperative Research Centre and the Chair of IEEE Task Force on Behavior and Social Informatics and of the IEEE Task Force on Educational Data Mining. He has served as an Associate Editor and Guest Editor on many journals. He has published 3 monographs, 4 edited books, 15 proceedings, 11 book chapters, and around 200 journal/conference publications. Dr. Cao is a Senior Member of the IEEE Systems, Man, and Cybernetics and Computer Societies.
\end{IEEEbiography}

\end{document}